\documentclass[11pt]{article}
\pdfoutput=1
\usepackage{jcappub,natbib,float,comment,bm,placeins} 

\def\be{\begin{equation}}
\def\ee{\end{equation}}
\def\ba#1\ea{\begin{align}#1\end{align}}

\renewcommand{\v}[1]{\mathbf{#1}}
\newcommand{\vx}{\v{x}}

\newcommand{\refeq}[1]{Eq.~(\ref{eq:#1})}          
          
\newcommand{\reffig}[1]{figure~\ref{fig:#1}} 

\newcommand{\refFig}[1]{Figure~\ref{fig:#1}}
          
\newcommand{\refsec}[1]{section~\ref{sec:#1}}          
\newcommand{\refSec}[1]{Section~\ref{sec:#1}}          
\newcommand{\refapp}[1]{Appendix~\ref{app:#1}}

\newcommand{\Om}{\Omega_m}
\newcommand{\Ob}{\Omega_b}

\renewcommand{\d}{\delta}

\def\iMpch{\,h\,{\rm Mpc}^{-1}}

\def\Mpch{\,h^{-1}{\rm Mpc}}

\newcommand{\bir}[1]{b_#1^{\rm rel}}

\newcommand{\LCDM}{\Lambda{\rm CDM}}

\title{The impact of massive neutrinos on halo assembly bias}

\author[a,b,c]{Titouan Lazeyras,}
\author[d]{Francisco Villaescusa-Navarro,}
\author[a,b,c,e]{Matteo Viel}
\affiliation[a]{SISSA, Via Bonomea 265, 34136 Trieste, Italy}
\affiliation[b]{INFN, Sezione di Trieste, Via Bonomea 265, 34136 Trieste, Italy}
\affiliation[c]{IFPU, Institute for Fundamental Physics of the Universe, via Beirut 2, 34151, Trieste, Italy}
\affiliation[d]{Department of Astrophysical Sciences, Princeton University, Peyton Hall, Princeton, NJ, 08544, USA}
\affiliation[e]{INAF-OATS, via Tiepolo 11, 34131 Trieste, Italy}
\emailAdd{tlazeyra@sissa.it, fvillaescusa@princeton.edu, viel@sissa.it}
\abstract{Using the publicly available Quijote simulations, we present the first measurements of the assembly bias of dark matter halos in N-body simulations which include massive neutrinos. We focus on the dependence of the linear bias $b_1$ on three halo properties: 1) concentration $c$, 2) spin $\lambda$, and 3) ellipticity $s$. Although these simulations cover a large volume, superior to any future surveys, we do not detect any effect of neutrinos on the relations $b_1(c)$, $b_1(\lambda)$ and $b_1(s)$ at fixed halo mass. We further study the dependence of halo properties and environment on neutrinos, finding these quantities to be impacted by neutrino masses at the same level as assembly bias. We find that the effect of neutrinos on spin and shape can be largely attributed to the change in the cold dark matter $\sigma_8$ in neutrinos simulations, which is not the case for concentration.}

\begin{document}
\maketitle
\flushbottom

\section{Introduction}
\label{sec:intro}

It is now well established that most of the observed tracers of the large-scale structure (LSS) of the Universe, such as galaxies, reside in dark matter halos. Hence the statistics of halos determine those of galaxies on large scales, making their distribution one of the key ingredients of the theoretical description of LSS. In the context of perturbation theory, the statistics of halos are written in terms of bias parameters multiplying operators constructed out of the matter density field (see \cite{Desjacques:2016} for a recent review). On sufficiently large scales, a linear relation between the halo density field $\d_h$ and the matter one $\d_m$ is enough to describe the clustering pattern of halos
\be
\d_h(\vx,\tau) = b_1(\tau) \d_m(\vx,\tau) + \dots\,,
\label{eq:localbias}
\ee
where $\delta_a$ is the density contrast of the considered field, defined as $\delta_a=\rho_a/\bar{\rho}_a-1$, $b_1$ is the linear bias of dark matter halos and the dots indicate that we only wrote the first term of the expansion.

This bias parameter was commonly thought to depend only on the redshift and mass of the considered halo population, implying that the clustering of dark matter halos is unaffected by the halo environment. In the last fifteen years however, several studies showed that such a model of halo biasing is too simplistic, and that halo bias depends on halo assembly history and other halo properties (see for instance \cite{Sheth:2004, gao:2005, Gao:2006, Wechsler:2005, Jing:2006, Croton:2006, Angulo:2007, Dalal:2008, Faltenbacher:2009, Lacerna:2012, Sunayama:2015, Paranjape:2016, Lazeyras:2016, Salcedo:2017, Mao:2017, Chue:2018, Sato-Polito:2018, Paco_18} and references therein). This phenomenon is now known as \textit{assembly bias}. Note that some papers refer to the dependence of halo bias on halo properties as secondary bias, reserving the term assembly bias for the dependence of halo bias on the assembly history (i.e. mass accretion rate or formation redshift). We do not make this difference here, and from now on, when we refer to assembly bias, we mean that the clustering of halos depends on properties other than halo mass and redshift.

Ref.~\cite{Sheth:2004} were the first to show that halo formation depends on the halo environment, which provided the first indirect evidence of assembly bias. Shortly after, \cite{gao:2005} presented the first direct measurements of this effect, showing that halos with a lower half-mass assembly redshift are less clustered than average, and vice versa. Since then, numerous works, using numerical simulations, have studied and found assembly bias as a function of various halo properties such as formation time, concentration, spin, shape, substructure content or mass accretion rate. One thing that quickly became clear is that possible correlations between two halo properties are not sufficient to explain the change in clustering observed with respect to these quantities \cite{Croton:2006,Lazeyras:2016, Mao:2017}. Since then, many papers studied the dependence of halo properties and linear bias as a function of their local environment and formation time (see \cite{Borzyszkowski:2016, Lee:2016, Musso:2017, Han:2018, Mansfield:2019, Ramakrishnan:2019} for recent works). While it is now clear that the halo environment, characterized by its position in the cosmic web and surrounding tidal field, plays a crucial role in assembly bias, a fully consistent picture explaining the physical mechanisms behind assembly bias at all mass, and which halo properties dominate it is still lacking to this day.

Assembly bias is not only studied in simulations. Indeed, several studies claimed to have observed \textit{galaxy} assembly bias (see for instance \cite{Yang:2005, Lacerna:2011, Tinker:2012, Wang:2013}). However, some of these claims were re-investigated by further studies without clear evidence for assembly bias \cite{Lin:2015}. More recently, \cite{Miyatake:2015, More:2016} presented the first evidence for assembly bias on galaxy cluster scales. This signal was however re-investigated by \cite{Zu:2016, Busch:2017,Sunayama:2019} who concluded that the signal found in \cite{Miyatake:2015, More:2016} was due to projection effects. Their results for assembly bias on cluster scales are consistent with zero. Excitement was revived very recently with \cite{Obuljen:2020} who claimed to have detected anisotropic assembly bias \cite{Obuljen:2019} in the BOSS DR12 data.

Since halo assembly bias has been shown to be a significant effect (up to 30\%), it is crucial to incorporate it correctly in the modeling of LSS for the analysis of current and future survey data, since a given galaxy sample could preferentially reside in halos with particular properties. This will be even more relevant if galaxy assembly bias is confirmed to be detected convincingly in observations. Notice that in theory the complete perturbative bias expansion automatically takes into account assembly bias if all necessary terms are included (see section~9.2 of \cite{Desjacques:2016}). Nevertheless, several efforts have recently been made to incorporate it in semi-analytical models such as (decorated) HOD  and abundance matching techniques (see \cite{Hearin:2015, Chaves-Montero:2015, Lehmann:2015, Zehavi:2017, Contreras:2018, Zehavi:2019, Contreras:2020, Xu:2020} and references therein). These methods have the advantage to allow to extract information from the non-perturbative regime, where perturbation theory fails.

LSS observables are also affected by the presence of massive neutrinos \citep{Lesgourgues_Pastor_2006, Villaescusa-Navarro_2015, Wong_2008, Saito_2008, Saito_2009, Blas_2014, Peloso_2015, Wong_2015, Biagetti_2014, Ichiki-Takada,  LoVerde_2014, LoVerde_2014b, LoVerde_2016, Massara_2014, Zhu_2013, Zhu_2014, Inman_2016, Massara_2015, Villaescusa-Navarro_2012, Villaescusa-Navarro_2011, Arka_2020a, Arka_2020b,Cora_2019, Massara_2020, Chang_2019, Arka_2019, Dvorkin_2019, Valcin_2019, Zhu_2020, Bollet_2020, Brinckmann_2019,Parimbelli_2020}, 
and future surveys such as Euclid or WFIRST are expected to put stringent constraint on their total mass. Present constraints set an upper limit at the 2$\sigma$ C.L. of 0.12 eV using either Planck data (including lensing) in combination with BAO \cite{Planck_2018} or Planck, BAO and the Lyman-$\alpha$ forest \cite{palanque15}. However, we stress that these limits are model dependent and strictly valid in a one-parameter extension (the neutrino mass) of a vanilla $\Lambda$CDM model, obtained by using Bayesian statistics.
Due to their large thermal velocities, neutrinos suppress the clustering of matter on small scales and leave a scale dependent imprint on the clustering on large-scales, as was recently measured in $b_1$ by \cite{Paco_14, Castorina:2013, Chiang:2018}. Furthermore, the effect of neutrinos is degenerated with other ones (such as $\sigma_8$ \citep{Paco_18b, Chang_19}) and hence it is important to model their impact correctly in order to achieve percent level precision cosmology in an unbiased way with future surveys. In this paper we investigate the impact of massive neutrinos on halo assembly bias in the Quijote simulations \cite{Quijote} which include massive neutrinos as light particles. These simulations are state-of-the-art ones and cover very large volumes which should allow us to detect any effect of neutrinos on assembly bias relevant for future surveys. 

This paper is organized as follows. In \refsec{sims} we present the simulation suite used for our study and the halo finding procedure. \refSec{assemblybias} describes our procedure to measure assembly bias, while our results are presented in \refsec{results}. We first focus on the impact of neutrinos on assembly bias in \refsec{assemblynu} before investigating their impact on halo property and environment in \refsec{haloprop}. We conclude in \refsec{concl}. The appendices present the importance of including neutrinos in the halo definition (\refapp{nuhalo}), additional results for other neutrino masses than those presented in the main text (\refapp{addnumass}), and the effect of varying $\sigma_8$ on $b_1$ (\refapp{sigma8}), which is relevant for interpreting our results as we will discuss later. 

\section{Simulations and halo finding}
\label{sec:sims} 

In this section, we briefly describe the characteristics of our simulation suite and the procedure used to identify halos and determine their properties.

Our results are based on the suite of Quijote simulations described in 
\cite{Quijote}. These gravity-only simulations have a very large combined total volume, and cover a wide range of different cosmologies (see table 1 in \cite{Quijote}). We use a subset of five cosmologies dubbed ``fiducial'', ``$M_\nu^+$'',``$M_\nu^{++}$'',``$M_\nu^{+++}$'', and ``$\sigma_8^+$'' in the nomenclature of \cite{Quijote}. The fiducial cosmology is a flat $\Lambda$CDM one with the following parameters: $\Om=0.3175$, $\Ob=0.049$, $h=0.6711$, $n_s=0.9624$, and $\sigma_8=0.834$. The volume of each simulation is $1~{\rm Gpc}^3/h^3$ sampled with $512^3$ particles, yielding a mass resolution $m_p=6.56\times10^{11} M_\odot/h$. We use 500 realizations, probing a combined volume of $500~({\rm Gpc}/h)^3$; simulations were initialized with second-order perturbation theory (2LPT) at $z_i=127$.  

The neutrino simulations adopt the same parameters but further implement massive neutrinos as light particles, using $512^3$ mass elements again. Crucially, the total matter density $\Om$ and $\sigma_8$ are kept fixed between the fiducial and neutrinos simulations, while the more relevant quantity might have been the CDM+baryon fluid ones as we discuss later. The three neutrino simulations model a sum of the neutrino masses equal to 0.1, 0.2, and 0.4 eV for the $M_\nu^+$,$M_\nu^{++}$, and $M_\nu^{+++}$ respectively. An additional difference between the neutrino simulations and the fiducial one is that neutrino ones are initialized with the Zel'dovich approximation. 

Finally, in order to disentangle\footnote{It is well known that the effect of massive neutrinos can be mimicked by a change in $\sigma_8$. Besides, when considering halos or galaxies, the relevant quantity is not the r.m.s. of the matter field on a sphere of $8~{\rm Mpc}/h$, $\sigma_{8,m}$, but of the CDM+baryons field $\sigma_{8,c}$ \citep{Paco_14, Castorina:2013}.} the effect of neutrinos to that of varying $\sigma_8$ in the CDM+baryons field ($\sigma_{8,c}$), we make use of the set of massless neutrino simulations with enhanced $\sigma_8$ which has a value $\sigma_8=0.849$. This value matches pretty well the value of $\sigma_8$ for the baryon-CDM fluid in the $M_\nu^{++}$ cosmology. These simulations are again initialized with 2LPT. We refer the reader to \cite{Quijote} for further detail on these simulations.

The dark matter halos were identified using the Amiga Halo Finder (hereafter AHF) \cite{Gill:2004,Knollmann:2009} at $z=0$, which identifies halos with a spherical overdensity (SO) algorithm. We set the value of the overdensity to 200 times the background density to define halos. By default, we take the cold dark matter-baryon fluid only to define halo (i.e. we do not include neutrinos in halos), but we do not expect the inclusion or not of neutrinos to affect our results in a significant way since they are much lighter than the rest. In addition, AHF removes unbound particles found within a halo, which we expect to be the case for a majority of neutrinos since, on these small scales, they are expected to have a large thermal velocity \citep{Paco_11, Paco_13}. However, we investigate the impact of including neutrinos in halos in \refapp{nuhalo}. 

The bound particles are then used to calculate canonical properties of halos like the density profile, rotation curve, mass, spin, and ellipticity. As this algorithm is based on particles, it natively supports multi-mass simulations. In this paper, we only use distinct halos and do not consider their sub-halos. Further, we will restrict our analysis to halos with at least 200 particles within $r_{200}$ (the radius corresponding to an overdensity of 200 with respect to the background density) to ensure convergence of the halo properties considered, such as concentration. With these parameters, the smallest halos have a mass of $10^{14.12} \, h^{-1} M_\odot$, well above $M_\star$ (the mass of halos with present day peak significance equal to 1). We have made use of 100 realizations of a simulation of the fiducial cosmology with a mass resolution eight times higher to conduct a convergence study of halo properties. We found the mean properties measured in the high resolution simulation to be within $3\%$ of the ones measured in the low resolution one at all mass, and already within $2\%$ for all but the lower mass bin we consider. Furthermore, in this paper we will study ratios between quantities measured in simulations including or not neutrinos, where convergence should be very robust. All of this justifies considering halos with down to 200 particles.


\section{Measuring assembly bias}
\label{sec:assemblybias}

We now describe the procedure we use to measure assembly bias. Even though neutrinos travel at high velocities, they are nonrelativistic at low redshift, and therefore they contribute to the total matter budget in the Universe, i.e.
\be 
\rho_m = \rho_{\rm cdm} + \rho_b + \rho_\nu,
\ee
where the subscript $b$ stands for baryons. Equivalently
\be 
\delta_m = f_{\rm cdm}\delta_{\rm cdm} + f_b\d_b + f_v\d_\nu = (1-f_\nu)\delta_c + f_\nu \delta_\nu,
\ee
where $f_i=\Omega_i/\Omega_m$ are the fractional background energy densities of each species, and in the last equality we assumed $\delta_b=\delta_{\rm cdm}\equiv\delta_c$ (which is correct for the simulations used in this paper) and used $f_b+f_{\rm cdm}=1-f_\nu$. 

The linear bias of dark matter halos can be obtained by taking the ratio of the halo-matter cross-power spectrum to the matter-matter auto-power spectrum, which tends to a constant, $b_1$, at low $k$. However, due to their large thermal velocities, that lead to large free streaming lengths $\lambda_{\rm fs} \gtrsim 100 \Mpch$, neutrinos are expected to create a scale-dependence in galaxy clustering on linear scales, as was measured in $b_1$ by \cite{Paco_14, Castorina:2013}. It was however shown in those works, that if halo bias is defined with respect to CDM+baryons only (subscript $c$ in the following), the bias becomes \textit{almost} scale-independent \citep{Chiang:2018,LoVerde_2014b}. This restores universality of the mass function and limits the scale dependence of halo bias on large-scales to a small effect (of the order $f_\nu \lesssim 1\%$). While this effect could also be present in assembly bias, we do not attempt to measure it here by lack of signal-to-noise ratio\footnote{Notice that \citep{Chiang:2018} used a very large set of N-body simulations, with a carefully set up for the neutrino component, to measure this effect. Given the realistic neutrino masses we are considering here, it is unlikely that we can detect this effect with our simulations.}, and we assume that the ratio of the power spectra indeed tends to a constant on the largest scales.
Hence, for a given halo sample of mass $M$, we define the large-scale linear bias as
\be
b_1(M)=\lim_{k\to 0} \frac{P_{\rm hc}(k,M)}{P_{\rm cc}(k)},
\label{eq:b1}
\ee
where the subscript $c$ stands for CDM+baryons. The extension of \refeq{b1} to other halo samples than halos selected by mass only is straightforward. In particular, if we further select halos in a given mass bin by a \textit{secondary} property $p$, we have
\be
b_1(p | M)=\lim_{k\to 0} \frac{P_{\rm hc}(k,p | M)}{P_{\rm cc}(k)}.
\label{eq:asblyb1}
\ee
In practice, and in order to maximize the signal-to-noise ratio, we make use of modes up to $k\sim 0.12 \iMpch$ by fitting a second order polynomial in order to take care of nonlinear effects appearing on smaller scales. We have checked the stability and robustness of our results under a change in this value and found it to be the best compromise in term of signal-to-noise ratio while still ensuring that our linear treatment is valid. This means that we measure the linear bias of a given halo sample as
\be
\frac{P_{\rm hc}(k)}{P_{\rm cc}(k)}=b_1+A k^2,
\label{eq:fit}
\ee
for $k<0.12 \iMpch$. The amplitude $A$ includes higher-order terms such as $b_2$ or $b_{K^2}$ in the bias expansion, as well as higher-derivative terms like $b_{\nabla ^2 \d}$ that we do not take into account in our modeling. Notice as well that because we use the cross-power spectrum in the numerator of \refeq{b1} we do not need to include a shot-noise term since it is negligible for dark matter particles.

We divide the total halo catalog into 5 tophat mass bins in logarithmic scale of equal size $\log M=0.26$ (where $\log$ is the base 10 logarithm) from $\log M = 14.12$ to $\log M = 15.43$. For each secondary property considered, we then divide the subcatalogs at each mass in 4 quartiles, and we measure the linear bias in each quartile using \refeq{asblyb1} to obtain the $b_1(p | M)$ relation.

In order to facilitate the comparison of our results with those of, e.g. \cite{Wechsler:2005,Lazeyras:2016}, we follow \cite{Wechsler:2005} and, for each halo property that we consider, we define
\be
p'= \frac{{\rm ln}(p/\bar{p})}{\sigma({\rm ln}p)},
\label{eq:prime}
\ee
where $p$ is the mean value of the property in a given quartile, $\bar{p}$ is the mean value in a given mass bin, and $\sigma$ is the square root of the variance at fixed mass. We then plot the relations $b_1(p' | M)$.

We investigate the dependence of the linear halo bias on three halo properties: the concentration $c$, spin parameter $\lambda$, and shape $s$. The halo concentration is quantified using the usual Navarro-Frenk-White (NFW) \cite{Navarro:1996} concentration parameter $c$ measured as in \cite{Prada:2011}. More specifically, AHF computes the ratio between the maximum of the circular velocity $V_{\rm max}$ and $V_{200}$, the circular velocity at $r_{200}$. For the case of the NFW halo profile \cite{Navarro:1996}, this ratio is given by
\be
\frac{V_{\rm max}}{V_{200}}=\left(\frac{0.216 \, c}{f(c)}\right)^{1/2},
\label{eq:cprada}
\ee
where $f(c)$ is given by
\be
f(c)=\ln (1+c)-\frac{c}{1+c}.
\label{eq:fofc}
\ee
Computing $c$ from the circular velocity at two different radii is hence straightforward. However, this way of inferring the concentration is not as robust as a proper fit of the halo density profile. For the halo spin, we use the spin parameter as defined in \cite{Bullock:2000}
\be
\lambda = \frac{|\v{J}|}{\sqrt{2}MVr_{200}},
\label{eq:spinparam}
\ee
where the angular momentum $\v{J}$, the mass $M$ and the circular velocity $V$ are evaluated at position $r_{200}$.
Finally, following the works of e.g. \cite{Faltenbacher:2009, Lazeyras:2016}, we also measure the bias as function of halo shape given by
\be
s=\frac{r_3}{r_1},
\label{eq:shape}
\ee
where $r_1 > r_2 > r_3$ are the axes of the moment-of-inertia tensor of the halo particles. This ratio is part of the standard output of AHF, which computes the moment-of-inertia tensor in the standard way from the distribution of particles in the sphere of radius $r_{200}$.

\section{Results}
\label{sec:results}

We now turn to our results. We start by measuring the impact of massive neutrinos on assembly bias, before turning to the dependence of halo properties and environment on neutrinos. We assess the importance of including neutrinos in the halo finding procedure in \refapp{nuhalo}. In this section we present results for the $M_\nu^{+++}$ cosmology (i.e. a neutrino mass of 0.4 eV) since we expect the potential effects to be maximal for this case. Results for neutrino masses of 0.1 and 0.2 eV are presented in \refapp{addnumass} for completeness.

\subsection{Impact of neutrinos on assembly bias}
\label{sec:assemblynu}

\begin{figure}
\centering
\includegraphics[scale=0.275]{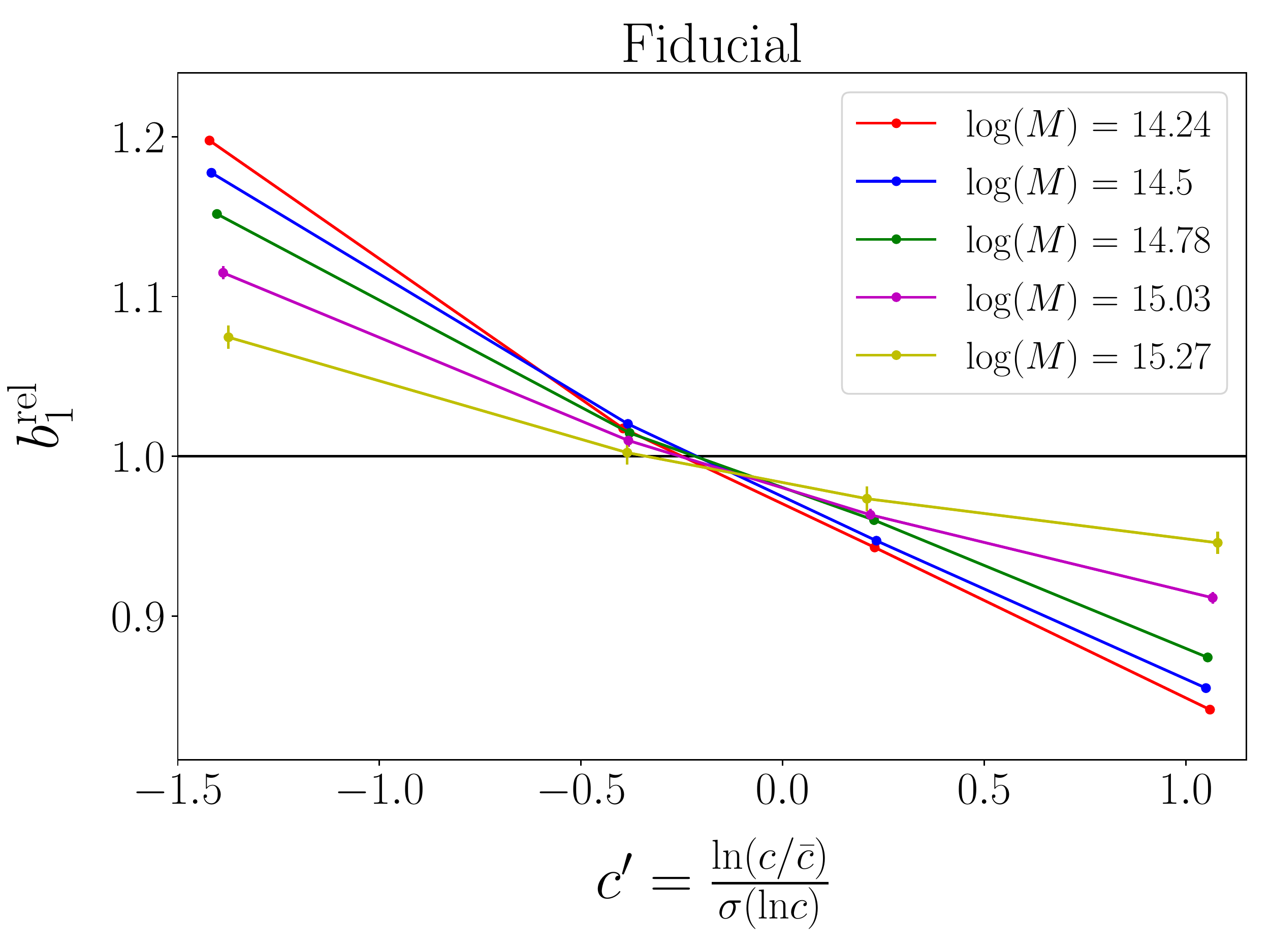}
\includegraphics[scale=0.275]{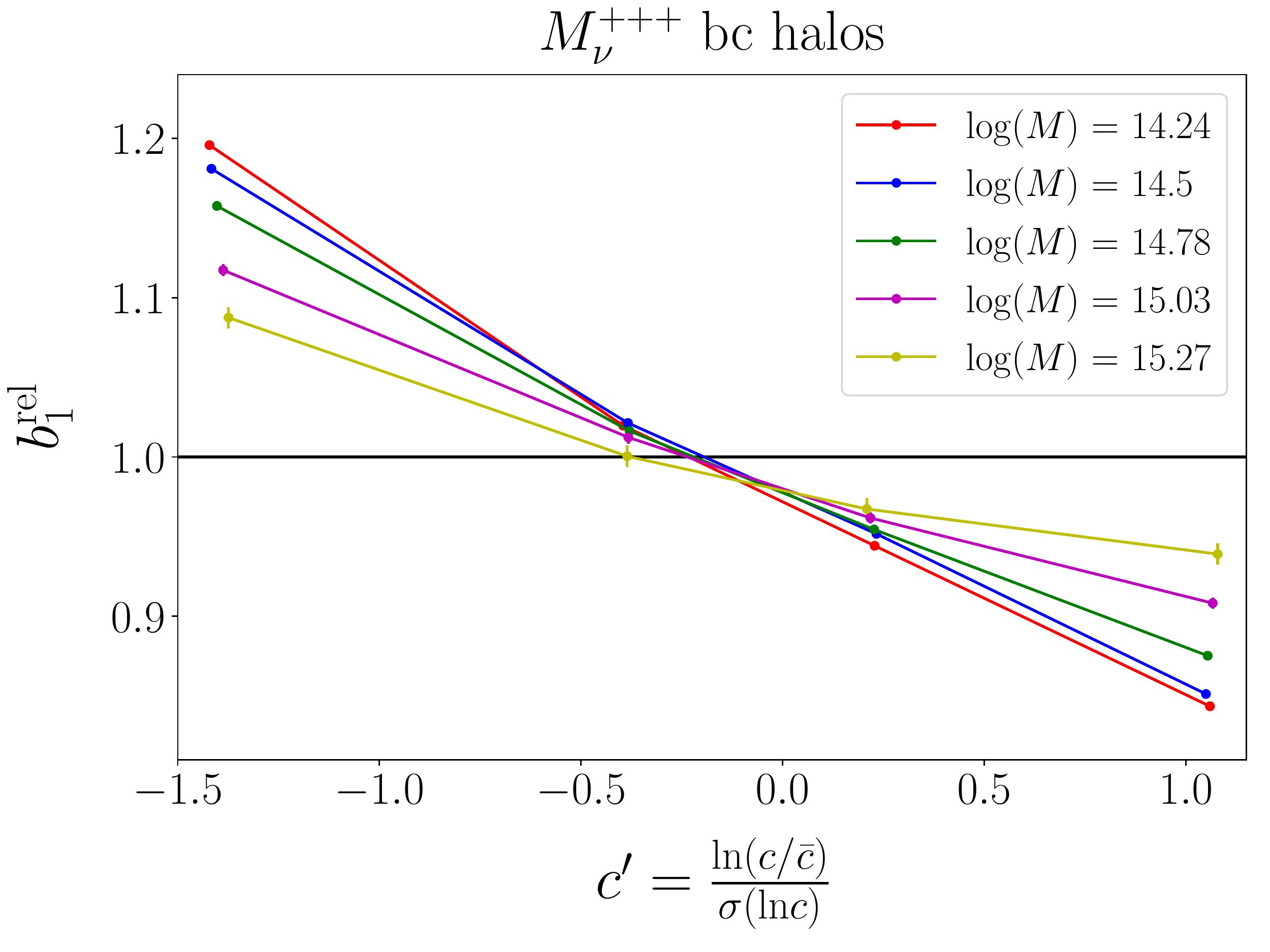}
\includegraphics[scale=0.275]{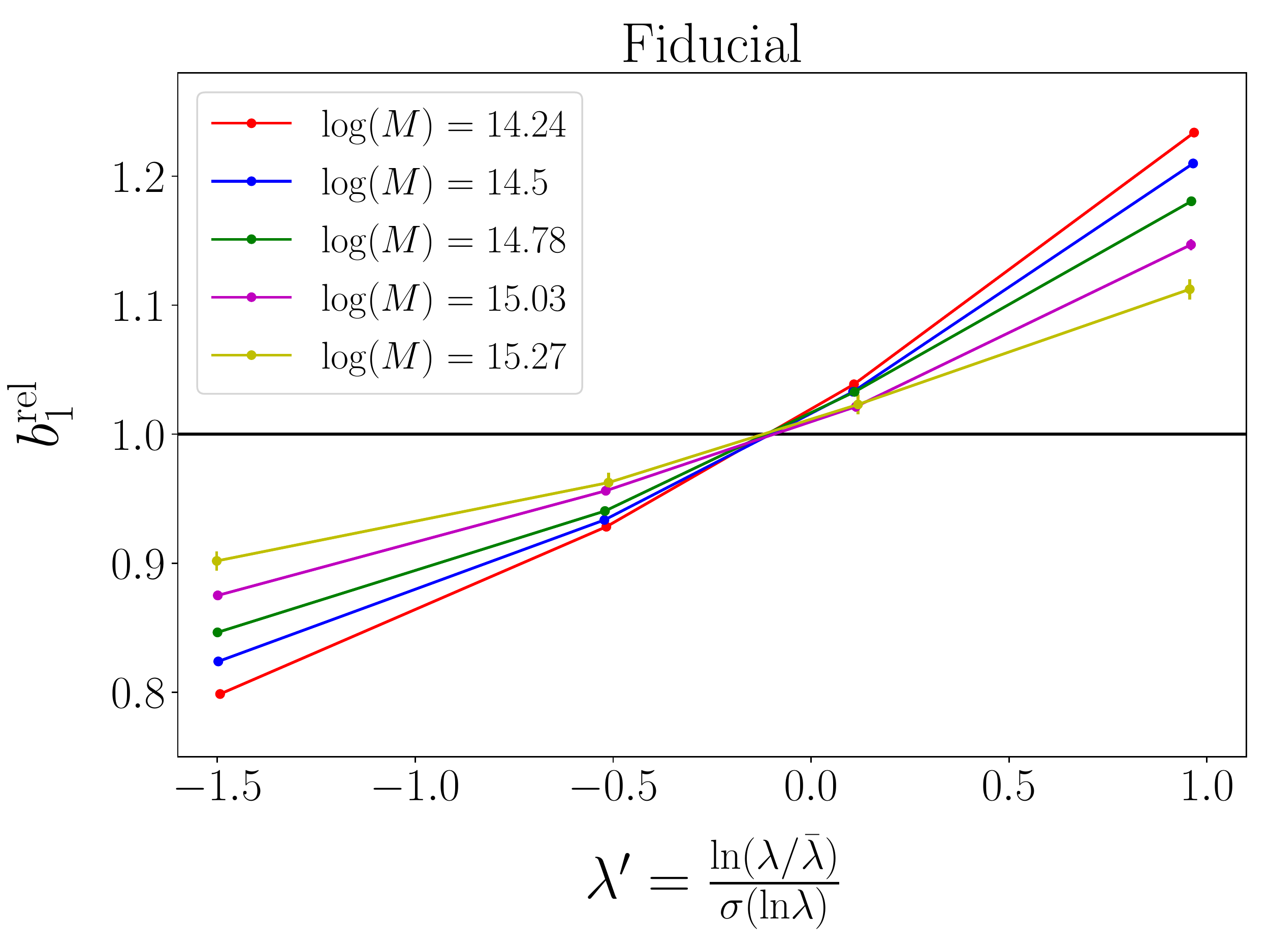}
\includegraphics[scale=0.275]{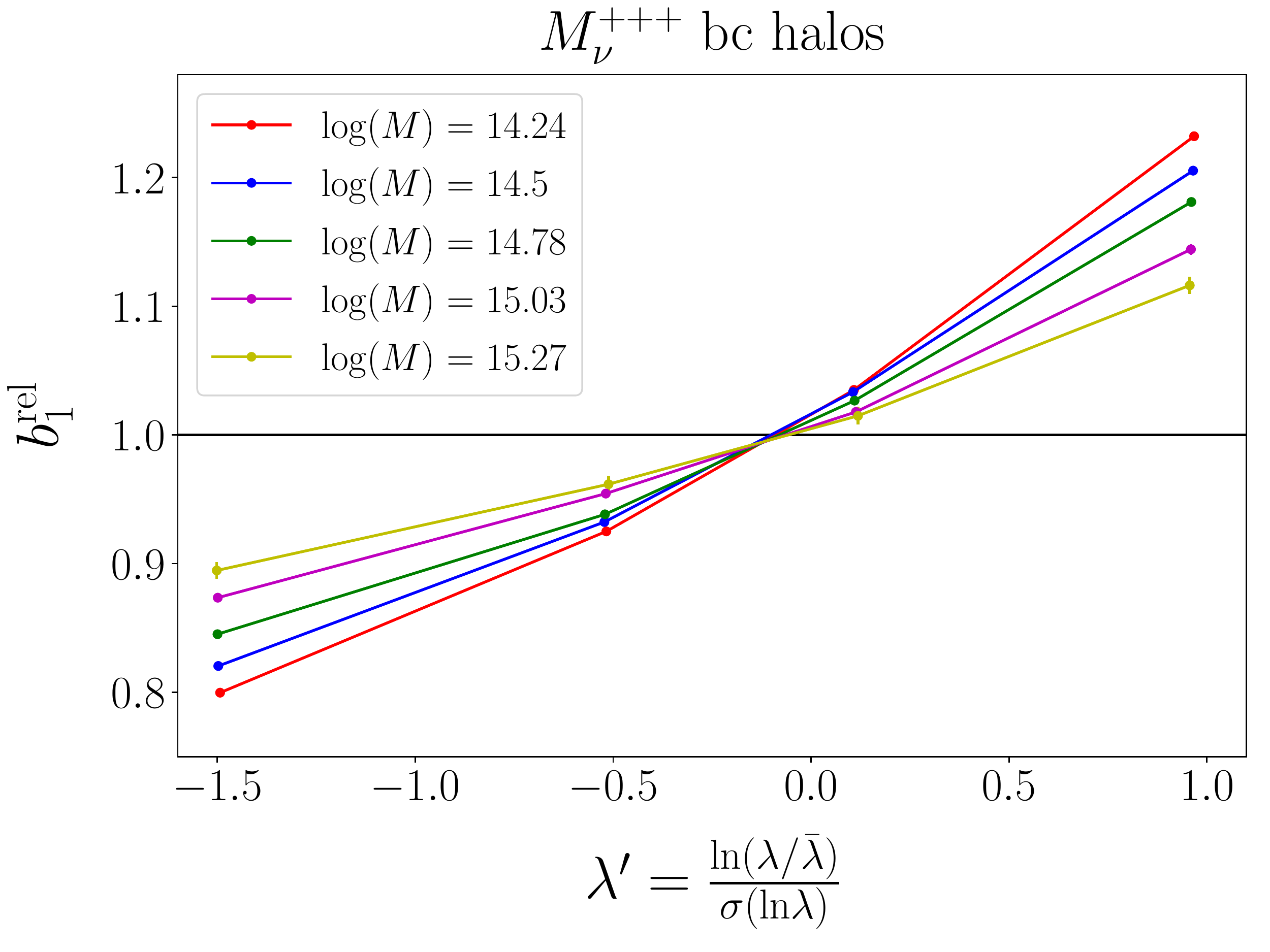}
\includegraphics[scale=0.275]{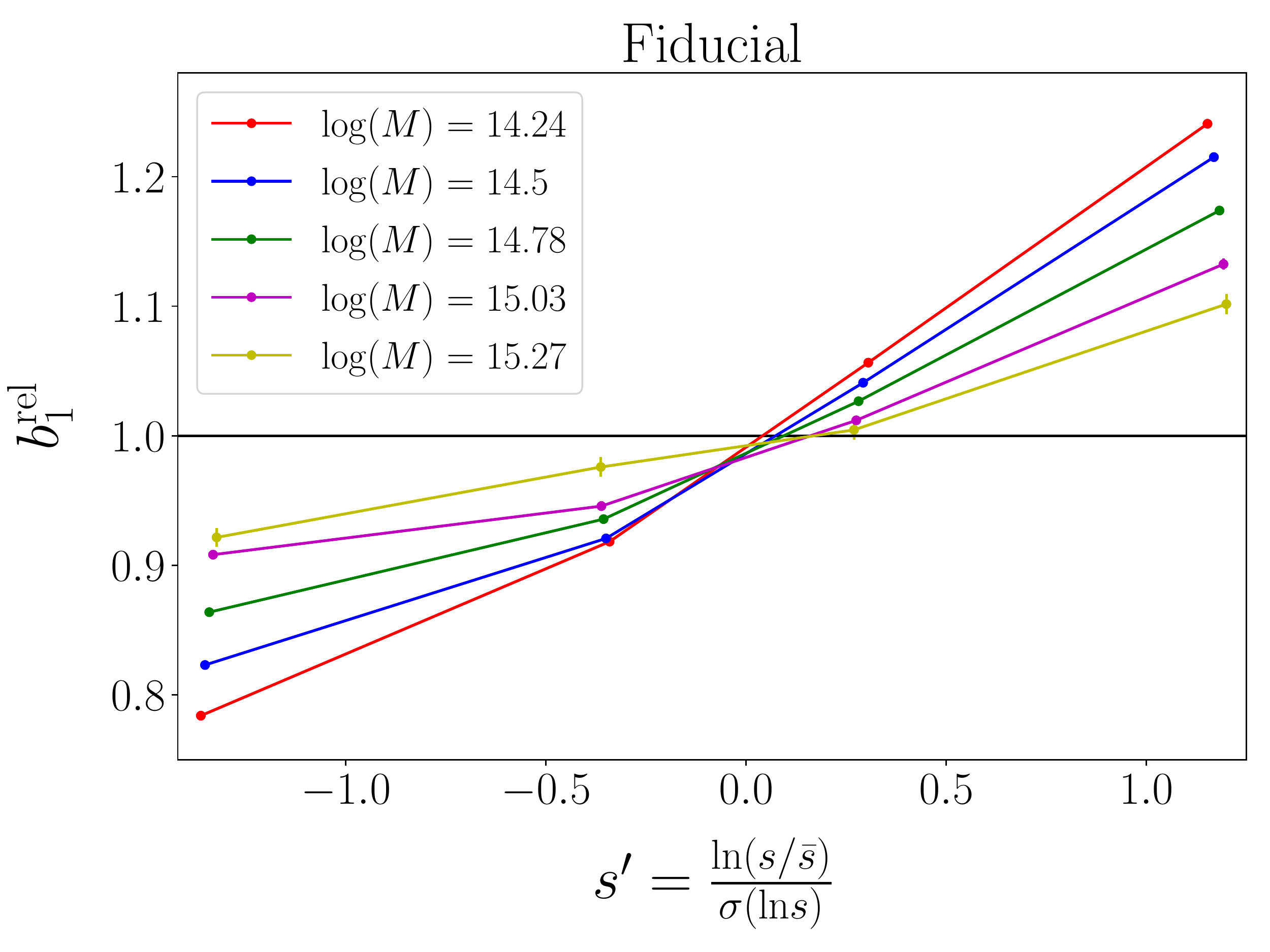}
\includegraphics[scale=0.275]{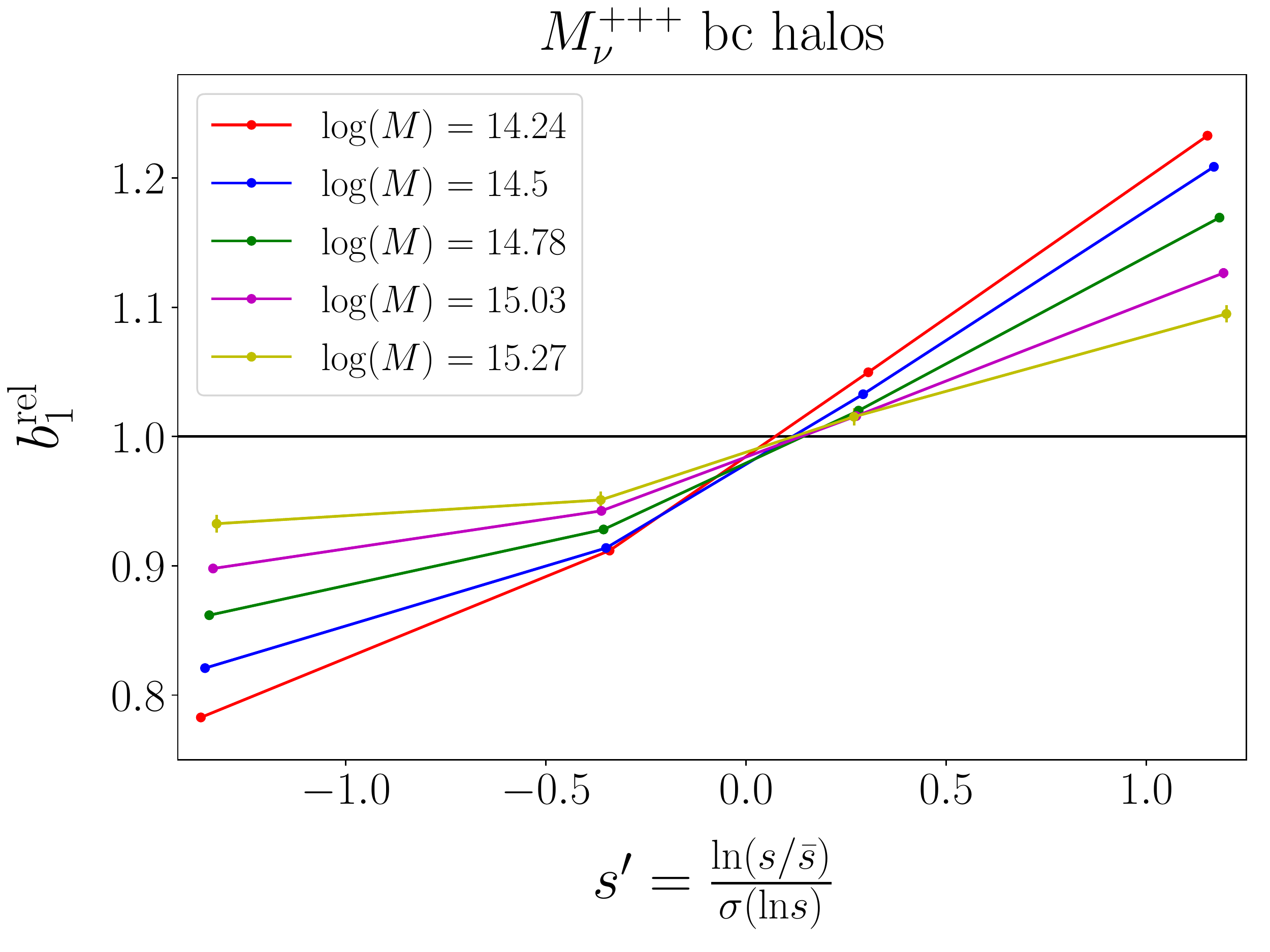}
\caption{Linear assembly bias $\bir{1}=b_1\left(p |M \right)/b_1(M)$ as a function of concentration (top row), spin (middle row), and ellipticity (bottom row) for several mass bins (indicated by the color coding). The errorbars show the $1\sigma$ error on the mean obtained from the 500 realizations. On each row, the left panel presents results for the fiducial cosmology, while the right one shows results for massive neutrinos simulations with a neutrino mass of 0.4 eV (``bc halos'' means that halos are obtained from the baryon-CDM fluid only). We recover all previously known trends in the fiducial case. No effect due to massive neutrinos can be detected on the right column.}
\label{fig:bofp}
\end{figure} 

\refFig{bofp} presents results for $b_1$  as a function of concentration $c$, spin parameter $\lambda$, and ellipticity $s$. The points linked by solid lines show the relative linear bias $\bir{1}=b_1\left(p |M \right)/b_1(M)$, and the errorbars show the $1\sigma$ error. We divide the bias as function of a property at fixed mass by the average one at the same mass to remove the vertical mass dependence that is of no interest in our context since it has already been extensively studied before. The color coding, indicating the mass, is the same on each figure and for each set of curves. Each time, the left panel presents results in the fiducial cosmology, while the right one shows results for the massive neutrinos simulations $M_\nu^{+++}$.

Before getting into the analysis of these figures, we would like to stress again that, although we will refer to low and high masses, one should keep in mind that all our results are for halo masses much above $M_\star$, and hence technically very massive halos (cluster-sized halos). 
Furthermore, as explained in \refsec{sims}, we used a spherical overdensity algorithm to identify halos. This most likely has an impact on our findings \cite{Chue:2018}, but we do not investigate how they would change if we used, e.g., a friends-of-friends (FoF) algorithm.

The results presented on \reffig{bofp} are in agreement with previous results from the literature (e.g. \cite{Gao:2006,Faltenbacher:2009, Paranjape:2016, Lazeyras:2016}). The top left panel of \reffig{bofp} shows the bias as a function of concentration (more concentrated halos are on the right) for the fiducial cosmology. We get a monotonically decreasing relation, and the effect is more important for the lower mass bins. The trends are very similar in all simulations involving massive neutrinos. On the other hand, the middle row of \reffig{bofp} shows that halos with a higher angular momentum are more clustered, and again the effect is stronger in the lower mass bins. Again, the effect of neutrinos is not visible by eye from these figures. Finally, the bottom row of \reffig{bofp} presents the linear bias as a function of halo shape at fixed mass. More spherical halos (with positive $s'$) are more biased, while more elliptical ones have a lower bias. The effect is larger for lower masses, and we cannot see any effect induced by neutrinos by eye.

We do not go in more details discussing these figures since all these results are already known and investigated in the case of massless neutrinos (e.g. \cite{Lazeyras:2016}), and that our goal here is to detect the effect of neutrinos on assembly bias.

\begin{figure}
\centering
\includegraphics[scale=0.275]{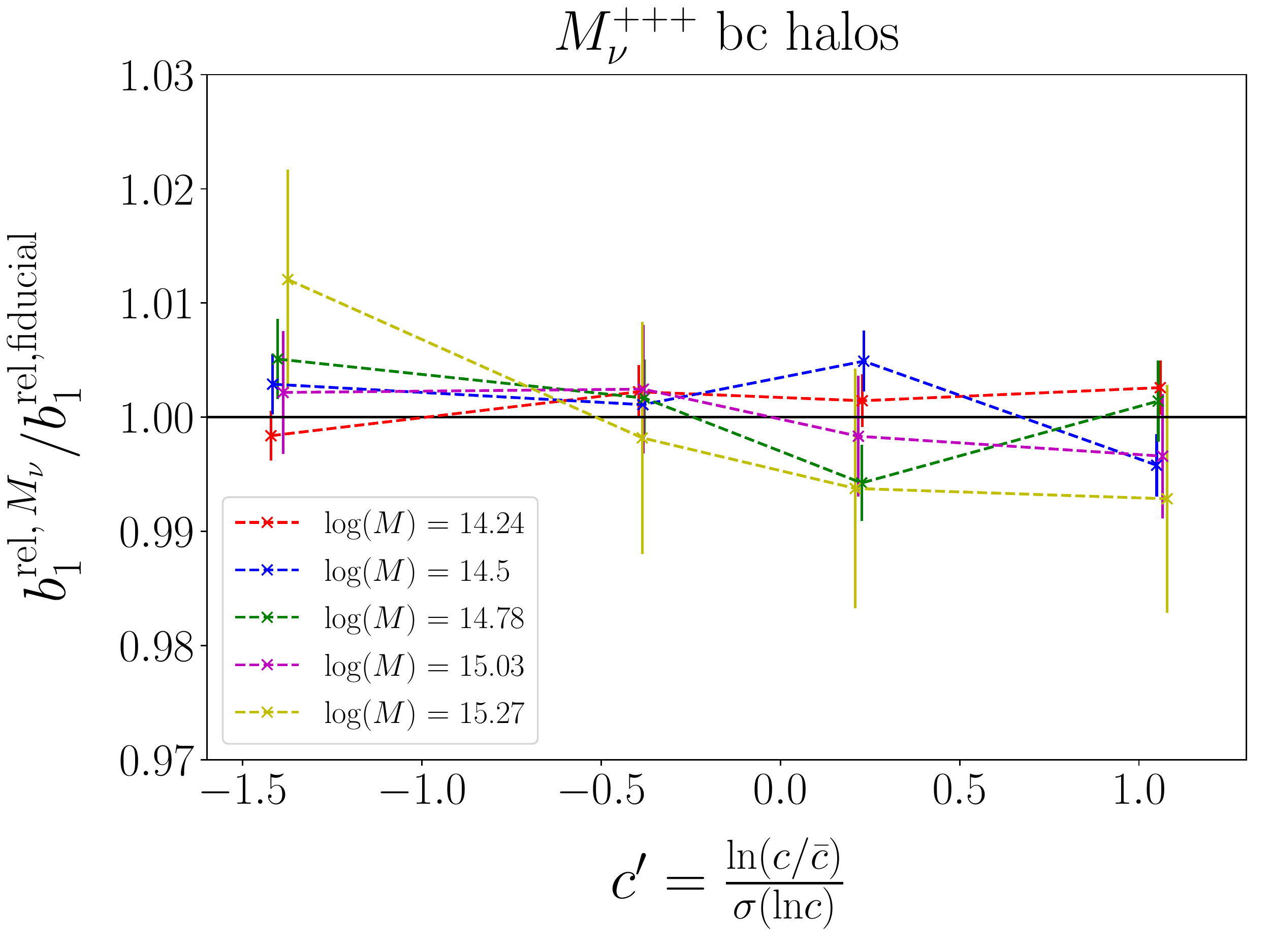}
\includegraphics[scale=0.275]{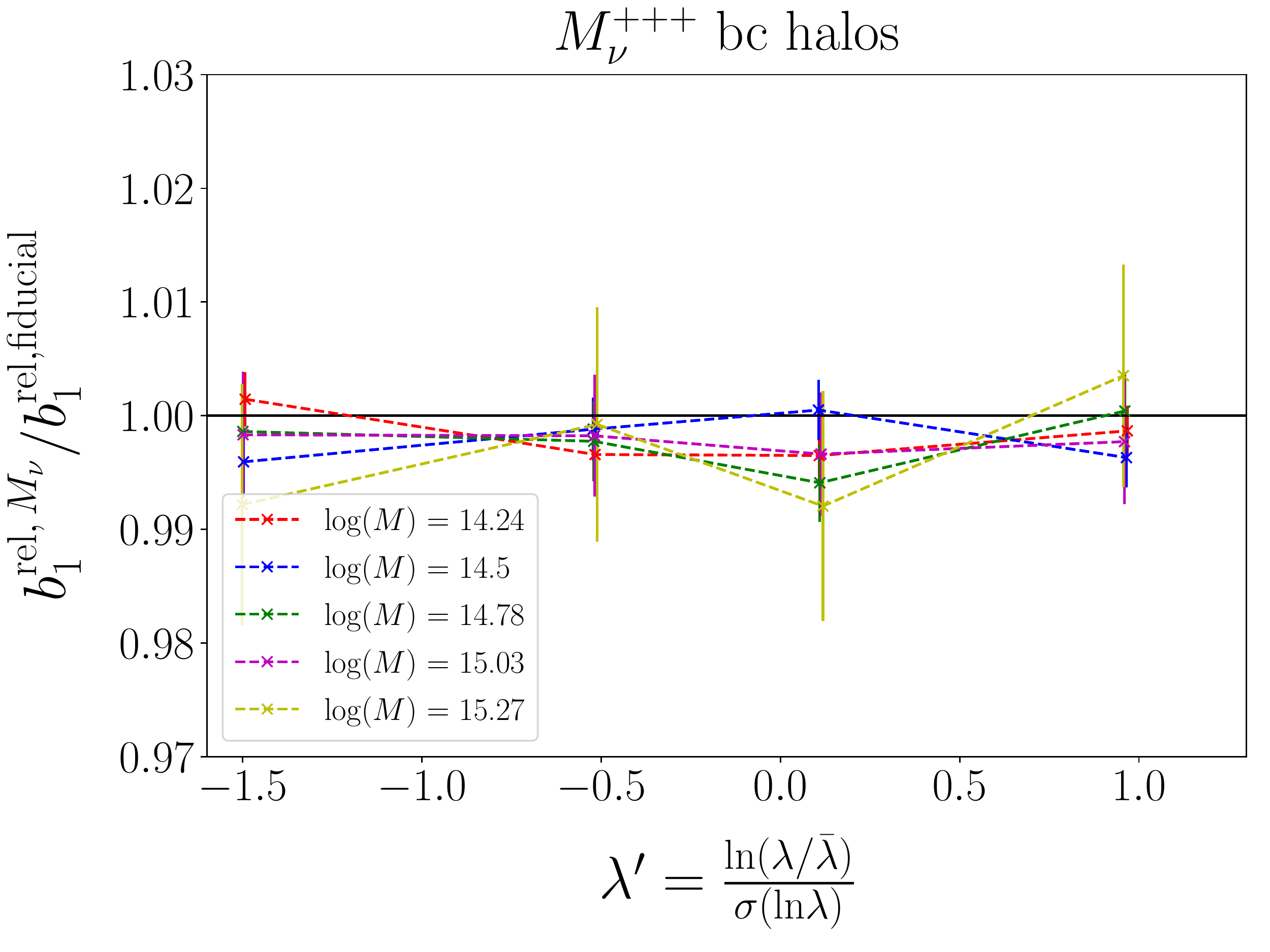}
\includegraphics[scale=0.275]{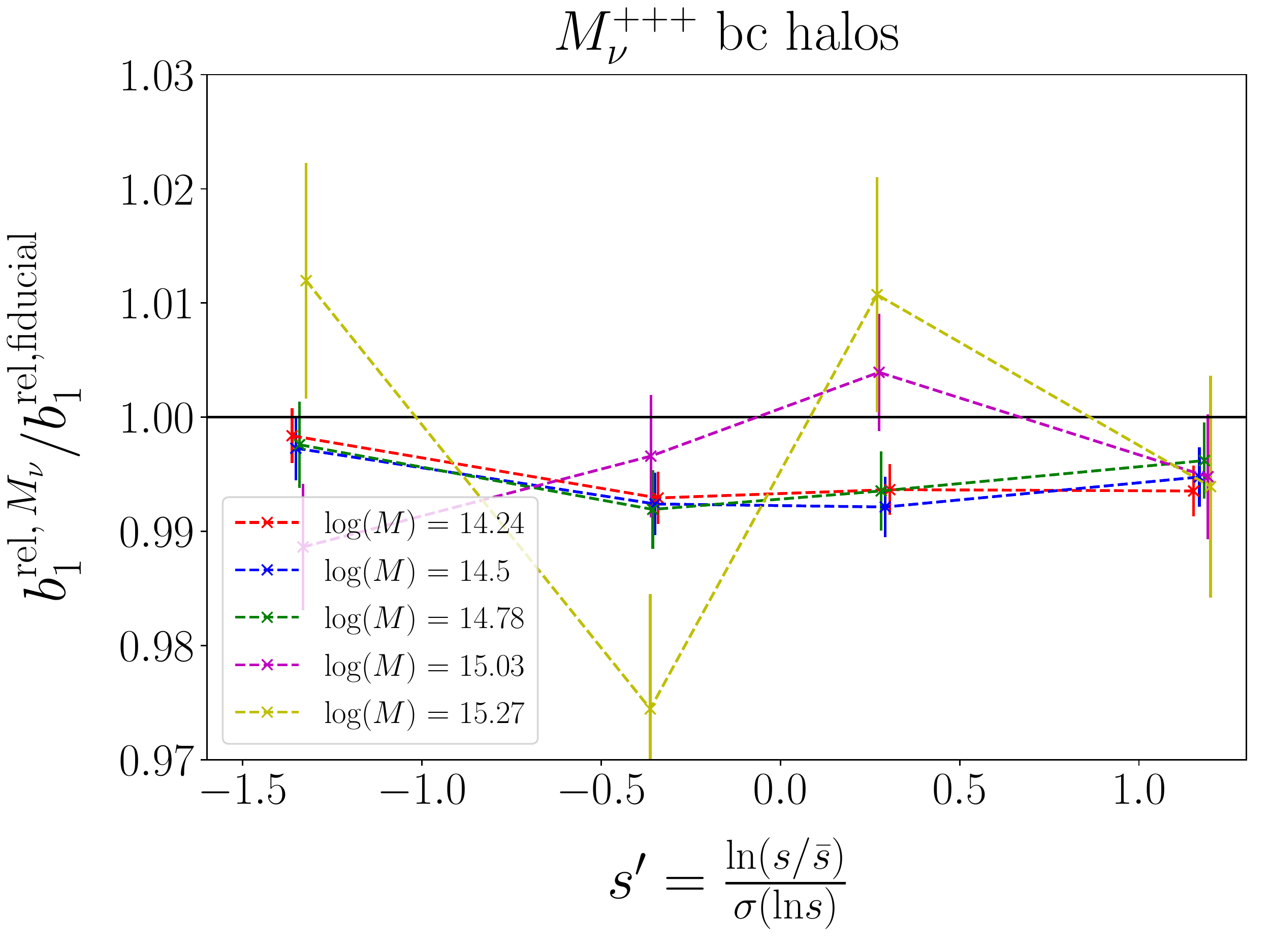}
\caption{Ratio of assembly bias in $b_1$ measured in simulations including neutrinos with a mass of 0.4 eV to the one measured in the fiducial cosmology as a function of concentration (top left), spin (top right), and ellipticity (bottom) for several mass bins (indicated by the color coding which is the same as the one on \reffig{bofp}.). Errorbars show the $1\sigma$ error on the mean obtained by a simple error propagation rule from the errorbars on \reffig{bofp}. Again, we do not include neutrinos in the halo finding (``bc halos''). On the bottom panel, more spherical halos have $s'$ closer to +1. As can be seen, no effect of neutrinos on the bias as a function of any of the properties can be detected with statistical significance.}
\label{fig:ratios}
\end{figure}

Since \reffig{bofp} does not allow to see any clear effect of the massive neutrinos on assembly bias, we further plot the ratios $b^{\rm rel, \, \nu}_1(p|M)/b^{\rm rel, \, fiducial}_1(p|M)$ (i.e. the ratio of the ratios), where the superscript ``$\nu$'' indicates that we measure this quantity in simulations including massive neutrinos, while the denominator is the one measured in the fiducial cosmology. The results are presented on \reffig{ratios} for the $M_\nu^{+++}$ simulations, and in \refapp{addnumass} for the two smaller neutrino masses. This figure is the main results of this paper. The color coding, indicating halo mass, is the same as on previous figures. The errorbars show the $1\sigma$ error on the mean obtained by error propagation from the errors showed on \reffig{bofp}.

We start again by discussing assembly bias as a function of concentration shown on the top left panel of \reffig{ratios}. More concentrated halos have a higher $c'$ value. We do not detect any measurable effect of neutrinos on the $b_1(c)$ relation at $M_\nu=0.4$ eV, and the effect is even smaller for masses of 0.1 and 0.2 eV. We could argue that there seems to be a mass-dependent, percent-level trend with more concentrated halos having a lower bias in neutrinos cosmologies. This is however not statistically significant given the size of the errorbars. We did not attempt to increase the signal-to-noise ratio by adding more simulation volume for two reasons: first, we would only be able to investigate the effect for the highest neutrino mass of 0.4 eV which is already an unreasonably high mass ruled out by observations \citep{Planck_2018}. Secondly we already use a total volume of $500~({\rm Gpc}/h)^3$ at a single redshift, which is much larger than any current or future survey volume.

Turning to $b_1(\lambda)$, the top right panel of \reffig{ratios} shows no clear effect due to neutrinos at any halo mass. The conclusions are the same at lower neutrino masses. One could argue a very small decrease in the overall amplitude of the bias (0.5\%) at all values of $\lambda$ in the simulation with total neutrino mass of 0.4 eV but this is again not statistically significant. Overall, we can safely conclude that neutrinos have no effect on assembly bias as a function of halo spin, at least for the volumes considered in this work.

Finally, the bottom panel of \reffig{ratios} shows the impact of neutrinos on the bias as a function of ellipticity in the $M_\nu^{+++}$ cosmology. More spherical halos are on the right with $s'$ closer to 1, while halos with a negative $s'$ are more elongated. As in the case of concentration and spin parameter, neutrinos do not seem to impact the relation $b_1(s)$. The only effect that can be seen is a percent effect on the overall amplitude of the bias at the three lowest mass considered in this paper. In this case there is a constant decrease of $b_1$ at all values of $s$. The two higher mass bins do not show this trend and have more scatter around the mean value, probably due to the smaller number of halos in these bins. Results presented on \reffig{ratios2} in \refapp{addnumass} show similar trends while less statistically significant with all curved consistent with 1.

The results presented on this figure imply that it will be very challenging to constraint neutrino masses from data using their impact on assembly bias\footnote{Notice that assembly bias in galaxies or galaxy clusters should in any case first be established.}. On the other hand, this also means that this is not an additional effect to take into account and model when analyzing future survey data. 

Before concluding this section we make a few important remarks. First, we note that we use only the baryon-CDM fluid in the halo finding procedure (``bc halos''), as explained before. While we do not expect that including neutrinos in halos will have a large impact, we inspect this in \refapp{nuhalo}. We indeed find the same results as in this section (\reffig{ratiownu}). Furthermore, we also check the impact of $\sigma_8$ on assembly bias in \refapp{sigma8}. This is important since, as we said before, the total matter $\sigma_8$ is kept fixed with respect to the fiducial cosmology in neutrino simulations. This means that the baryon-CDM fluid has a modified (enhanced) $\sigma_8$ which impacts halo clustering and can impact assembly bias. As shown in \reffig{ratiosigma8} we do not find any significant impact of $\sigma_8$ on assembly bias as a function of any properties. However we can observe some small trends by eye which go in the same way as the ones due to neutrinos discussed in this section. This minimizes the impact of neutrinos on assembly bias even further. Finally, using the 100 realizations of the high resolution simulation, we checked the impact of mass resolution on our results. We did not find any significant impact, and conclude that mass resolution does not induce any spurious assembly bias signal. Finally, we emphasis that choosing a more conservative minimum number of particles per halo such as 400 would only remove the lowest mass bin, leaving our conclusions that massive neutrinos do not impact the assembly bias of cluster-sized halos unchanged. 

\subsection{Dependence of halo properties and environment on neutrinos}
\label{sec:haloprop}

In this section we study the impact of neutrinos on halo properties, as well as their large-scale environment, defined as the overdensity (both of total matter and of baryon-CDM only) in a sphere of radius $R=20 \Mpch$.

\begin{figure}
\centering
\includegraphics[scale=0.275]{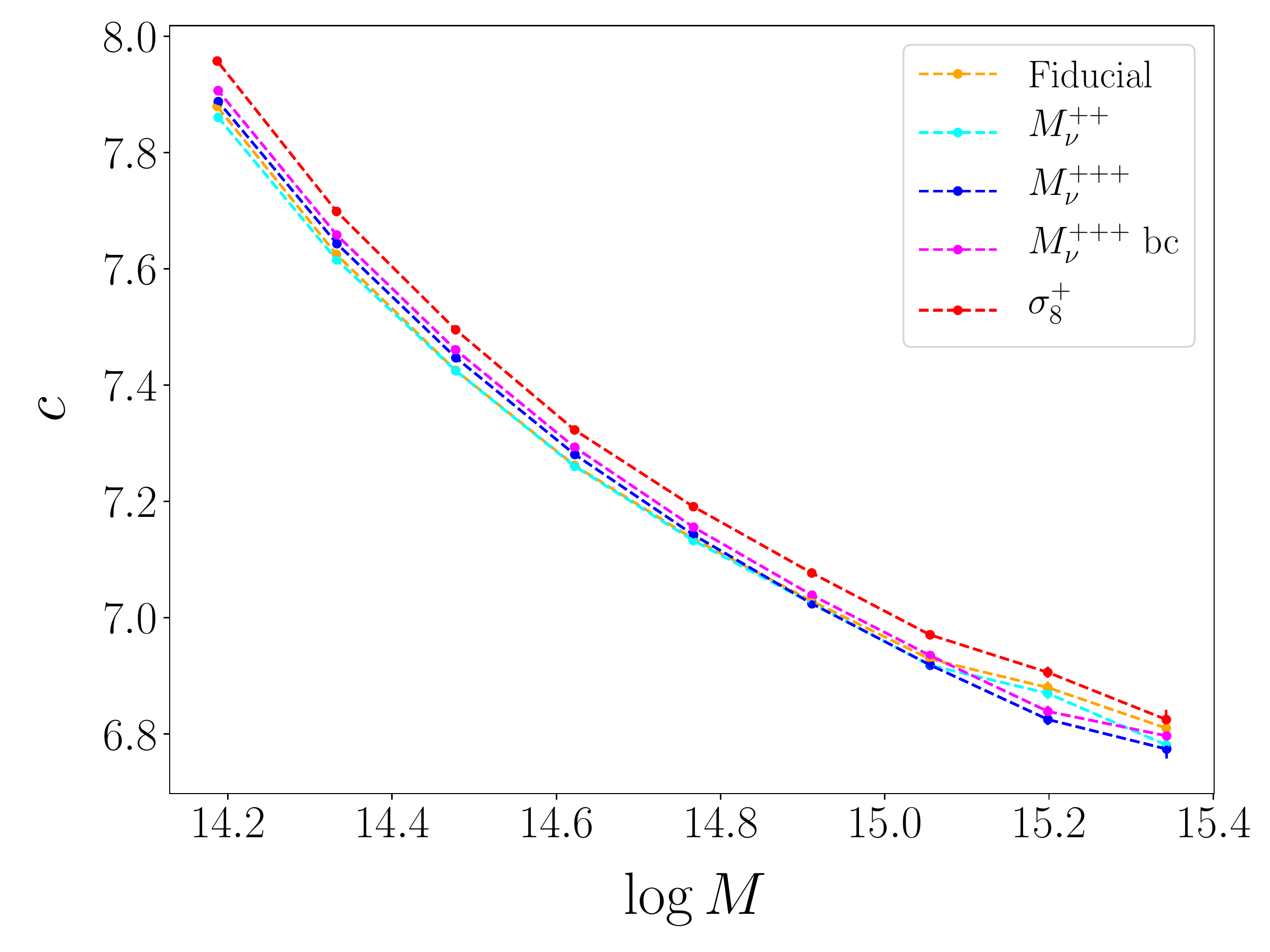}
\includegraphics[scale=0.275]{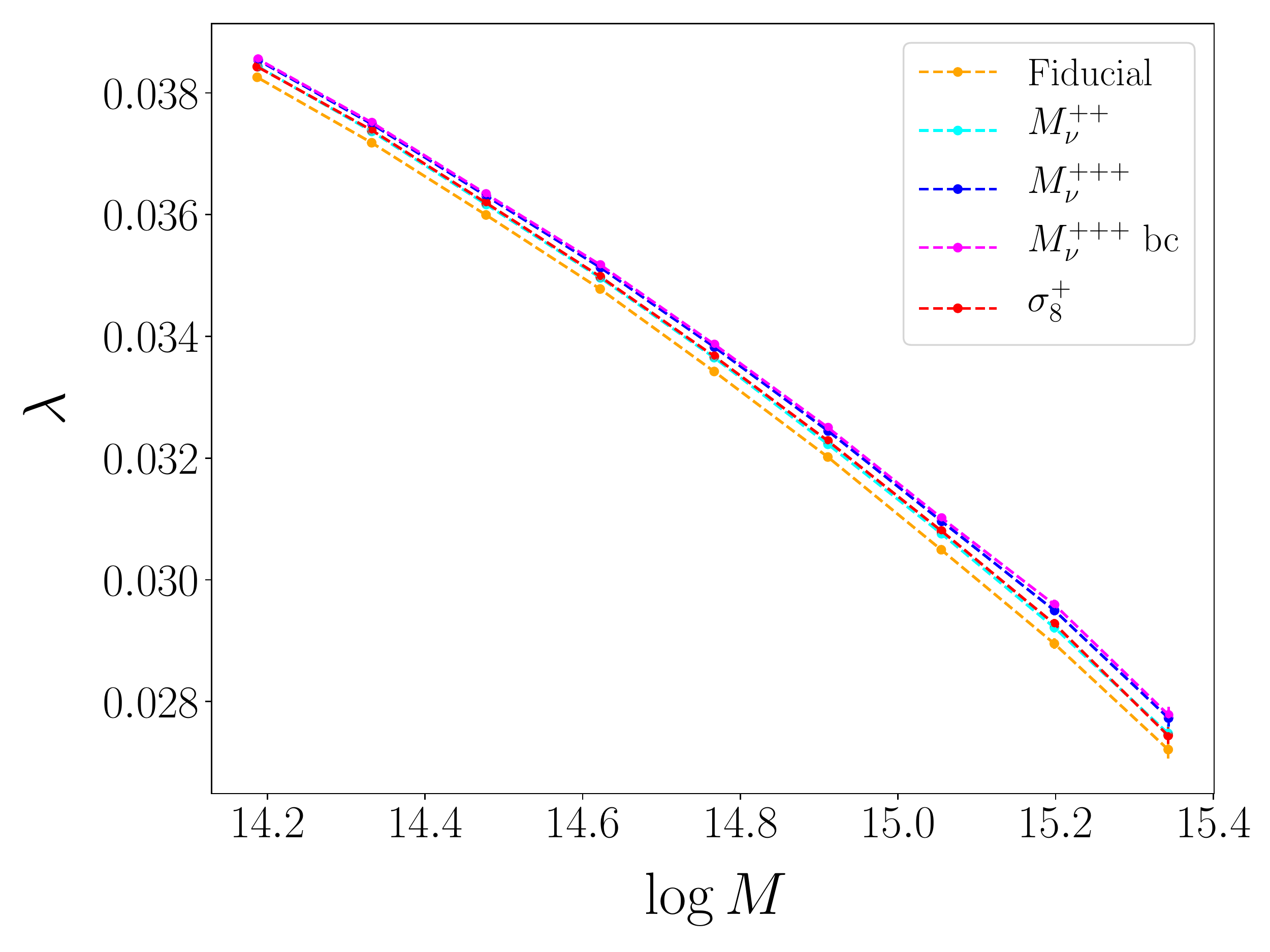}
\includegraphics[scale=0.275]{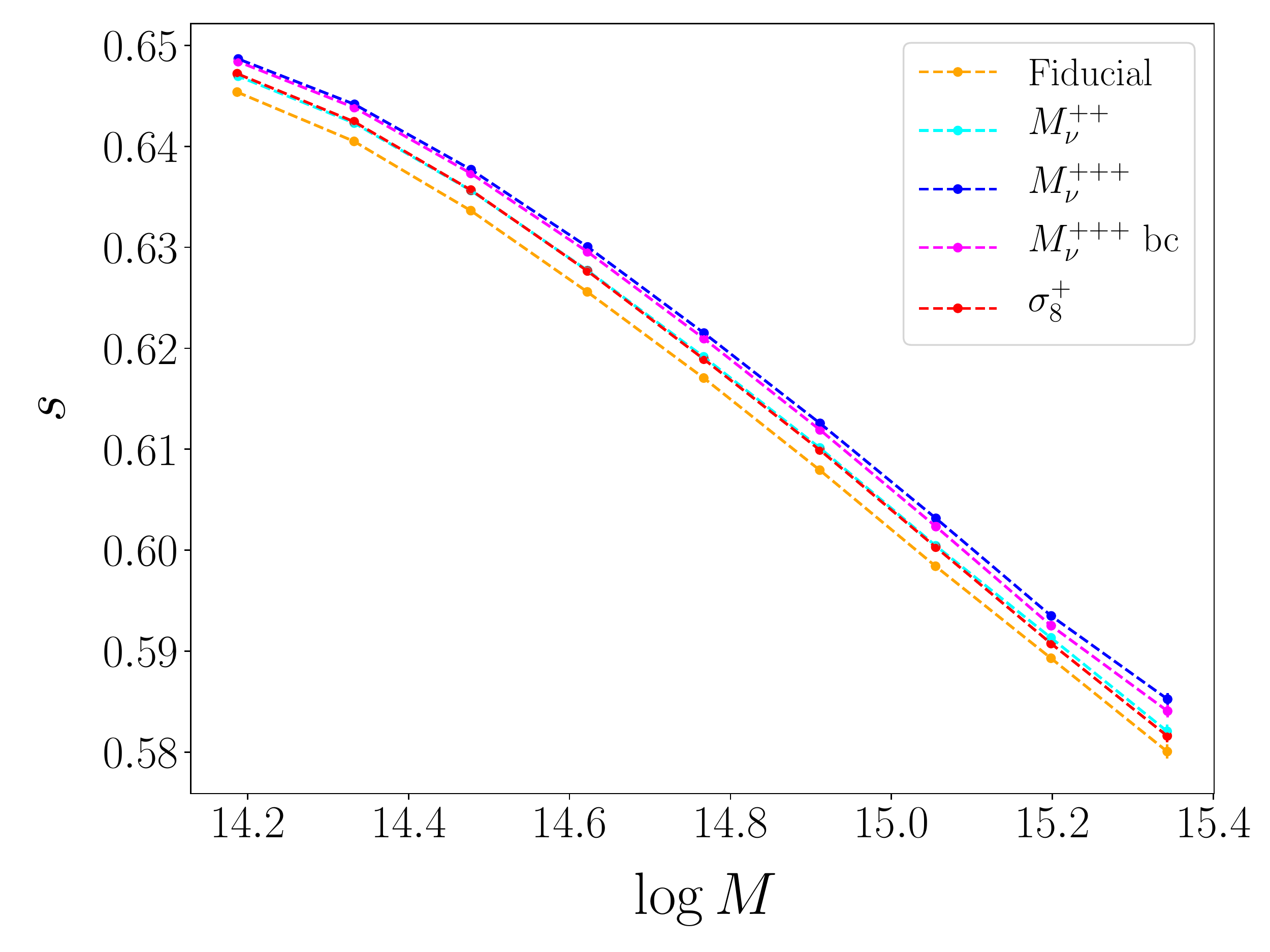}
\caption{The concentration (top left), spin (top right), and shape (bottom) of halos as a function of mass for different cosmologies. The relations showed are mean relations but of course the scatter (not shown) around these means is significant. Each time we show the relation in the fiducial case (orange) and in simulations with neutrinos of mass $0.2$ (cyan) and $0.4$ eV (blue) for halos built from the baryon-CDM+neutrino fluid. We further show the same relation for halos built only with the baryon-CDM fluid in the $M_\nu^{+++}$ cosmology in magenta to assess the importance of including neutrinos in halos. Finally we also show the effect of enhancing $\sigma_8$ in red ($\sigma_8^+$) in order to disentangle the effect of $\sigma_8$ from the one of neutrinos. The general trends common to all cosmologies are in agreement with previous works. To our knowledge this is the first time that the effect of neutrinos on these properties is studied. However we see that the effect of neutrinos on $\lambda$ and $s$ can be largely attributed to the change in the baryon-CDM $\sigma_8$ while this is not the case for $c$ where the effect of neutrinos and $\sigma_8$ seem to counteract each other. See text for more details.}
\label{fig:propsofM}
\end{figure}

We start with the concentration, spin, and ellipticity - mass relations in order to investigate how neutrinos impact them. Since assembly bias is defined as the dependence of halo bias on properties other than mass, it is logical to assess the level at which neutrinos affect dark matter halo properties. We note that we do not expect any of the studied quantities to be impacted by neutrinos significantly, (i.e. more than assembly bias itself) since otherwise this would lead to a spurious signal in assembly bias due to an Eddington-like bias in the case of mass. To be clearer, if neutrinos affected strongly a given property $p$ without affecting the clustering itself (i.e halos would be assigned a ``wrong'' value for $p$), assembly bias as a function of $p$ would then be different in the fiducial and neutrino cosmologies, which in turn would show on \reffig{ratios}. This is however not what we observed, which is why we do not expect any strong impact of neutrinos on the three halo properties studied in this paper.

The results are shown in \reffig{propsofM} for the fiducial case (orange) and in simulations with neutrinos of mass $0.2$ (cyan) and $0.4$ eV (blue) for halos built from the baryon-CDM+neutrino fluid. We further show the same relation for halos built only with the baryon-CDM fluid in the $M_\nu^{+++}$ cosmology in magenta to assess the importance of including neutrinos in halos. Finally we also show the effect of enhancing $\sigma_8$ in red ($\sigma_8^+$) in order to disentangle the effect of $\sigma_8$ from the one of neutrinos. The relations showed are mean relations but of course the scatter (not shown) around these means is big. We do not go into a detailed analysis of the general trends common to all cosmologies in these relations but simply note that we recover all the already known ones (e.g. \cite{Ludlow:2016, Despali:2016, Child:2018} and references therein). In particular more massive halos tend to be less concentrated, have a smaller angular momentum and are more elliptical (this last relation can seem a little surprising but can be understood, for very massive objects, when realizing that more spherical halos are also more concentrated and hence correspond to less massive objects, e.g. \cite{Lazeyras:2016}). We see that neutrinos have a small effect on all three properties, and that including them or not in the halo definition has no effect on the inference of properties. We develop further on this last point and its impact on assembly bias in \refapp{nuhalo}. A final point to note is that, while the impact of enhancing $\sigma_8$ is the same as the one due to neutrinos in the case of $M_\nu=0.2$ eV for the spin and shape of halos, it is not the case for concentration. Indeed enhancing $\sigma_8$ increases halo concentration but neutrinos tend to slightly make it smaller, as expected. We hence conclude that the effect of neutrinos on spin and shape can be largely attributed to the change in $\sigma_8$ but not for the case of concentration.

\begin{figure}
\centering
\includegraphics[scale=0.275]{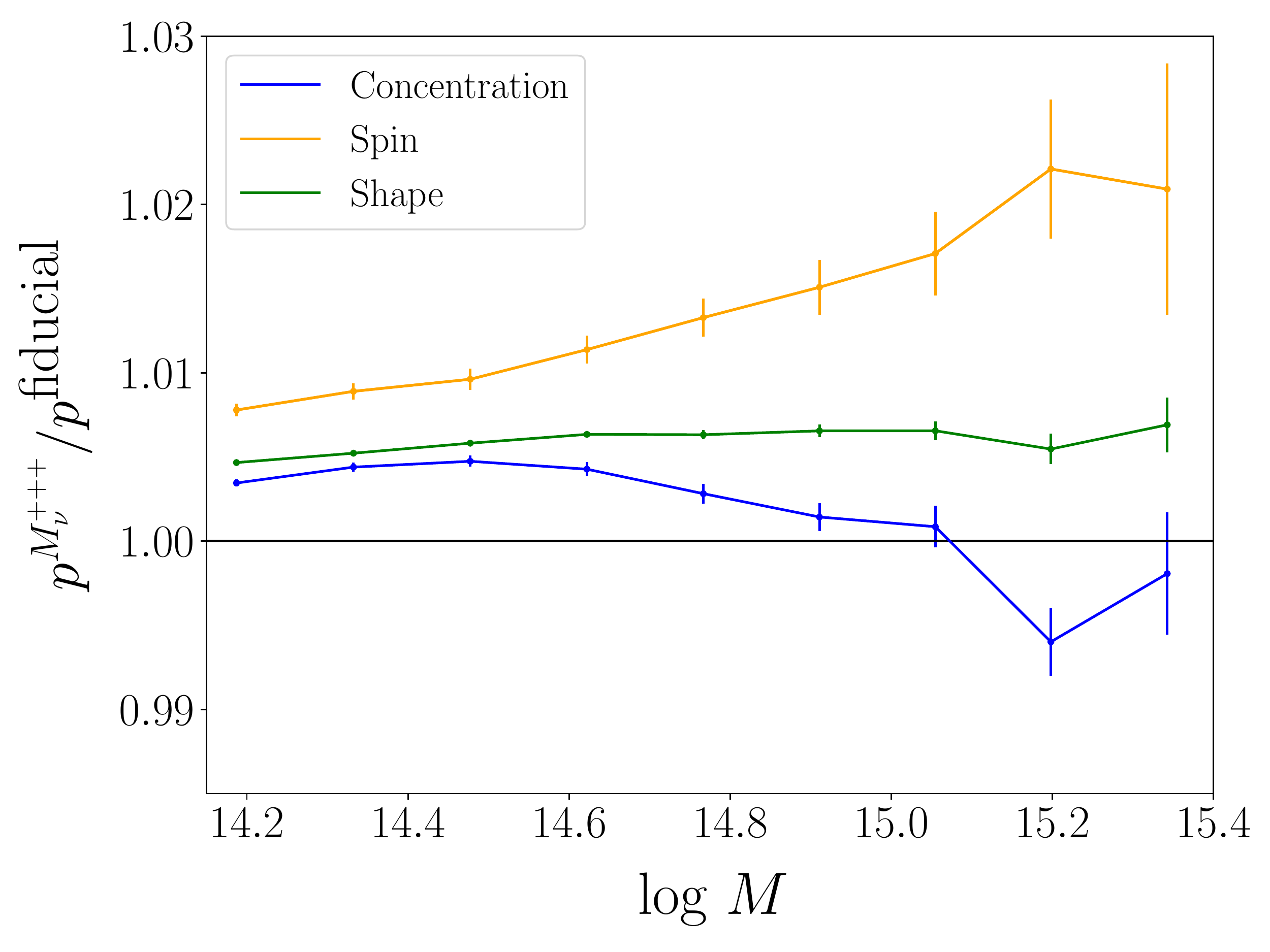}
\includegraphics[scale=0.275]{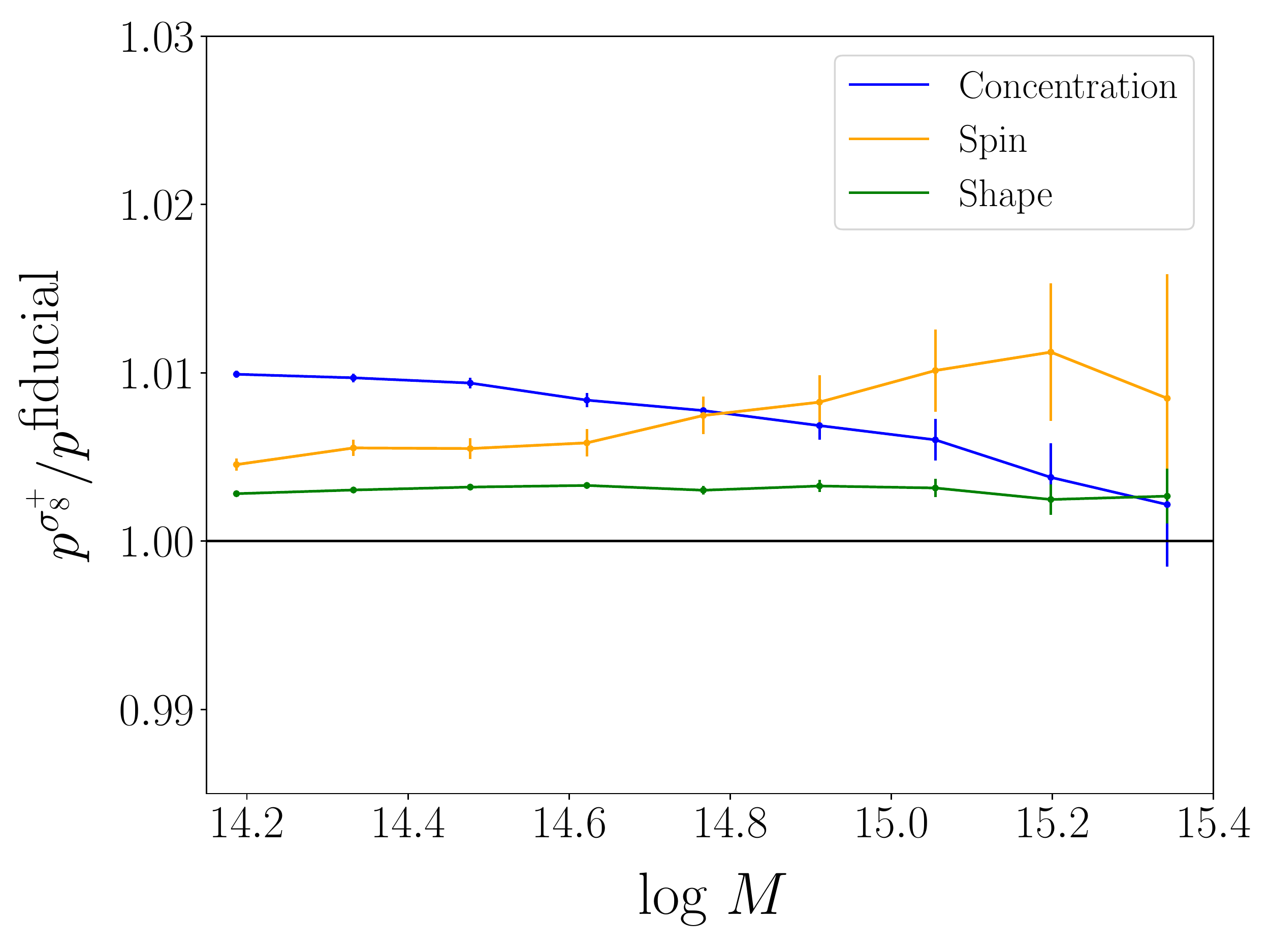}
\caption{Ratio of each property in the $M_\nu^{+++}$ (left) and $\sigma_8^+$ (right) cosmology to the one in the fiducial simulation. The errorbars show the $1\sigma$ error on the mean obtained by error propagation from the errors of the $c/L/s(M)$ relations. We see that the effect of neutrinos is of a couple percents maximum, and that halo spin is the most affected. We however see that the effect of neutrinos on $\lambda$ and $s$ seems to be largely attributed to the change in the baryon-CDM $\sigma_8$ (with smaller amplitude since the change in $\sigma_8$ presented here corresponds to a neutrino mass of 0.2 eV), while this is not the case for $c$ where the effect of neutrinos and $\sigma_8$ seem counteract each other. See text for more details.}
\label{fig:ratiopropsofM}
\end{figure}

In order to be more quantitative, \reffig{ratiopropsofM} shows the ratio of each property in the $M_\nu^{+++}$ cosmology to the one in the fiducial simulation (left panel), as well as for the $\sigma_8^+$ case (right panel). Clearly neutrinos affect all quantities in a mass dependent way, the largest effect being seen at large masses, and that effect is of a couple percents maximum. We see again that the change in $\sigma_8$ seems to explain the effect of neutrinos on $\lambda$ and $s$ (with smaller amplitude since the change in $\sigma_8$ presented here corresponds to a neutrino mass of 0.2 eV), but not on $c$. These results go in the same direction as what we observed for assembly bias and are hence expected for reasons that we explained before. More specifically, halos in massive neutrinos cosmologies tend to be slightly more spherical and have a larger spin parameter, and their concentration tends to decrease at very high mass. While the effect on concentration has been partially studied before \citep{Brandbyge_2010,Paco_13}, it is to our knowledge the first time that  the dependence of halo spin and ellipticity on neutrino mass is measured (see however \cite{Lee:2020} for the effect of massive neutrinos on halo spin flip). These results can be easily understood when considering the large velocity dispersion of neutrinos (in particular this directly explains the larger angular momentum observed) but, as we explained before, they can be largely attributed to the change in the baryon-CDM $\sigma_8$. A maybe more surprising result is the fact that the concentration is only very weakly affected by neutrinos, since one would expect the large free streaming scale of neutrinos to diminish the clustering of matter and hence also the concentration of halos by making their profile more shallow. This might still be true. However, since $\sigma_8$ for the baryon+CDM fluid also varies in neutrino cosmology and tends to increase the concentration, the two effects probably oppose and cancel each other leading to no net total difference between halo concentration in the fiducial and neutrino cosmologies.

\begin{figure}
\centering
\includegraphics[scale=0.275]{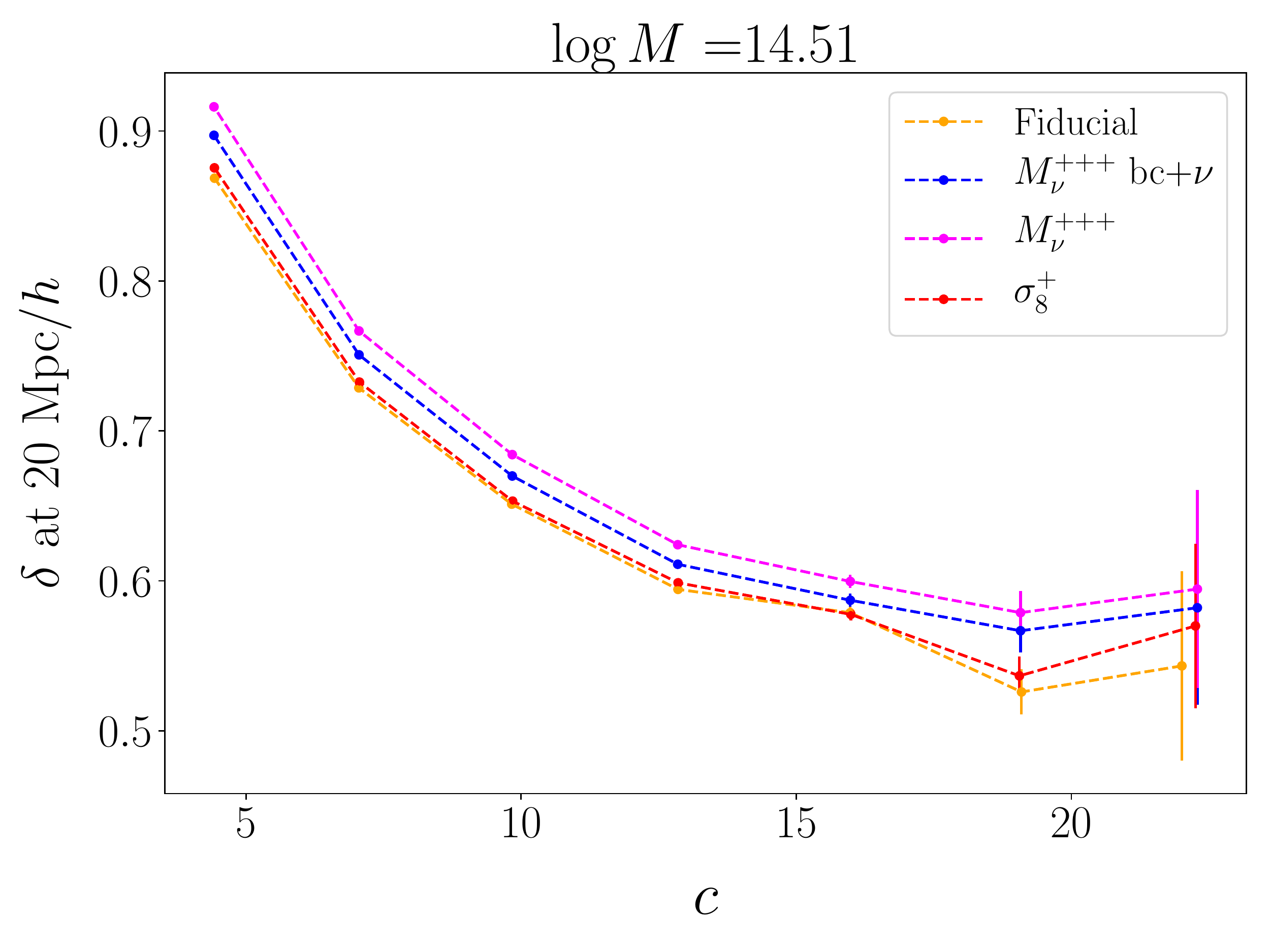}
\includegraphics[scale=0.275]{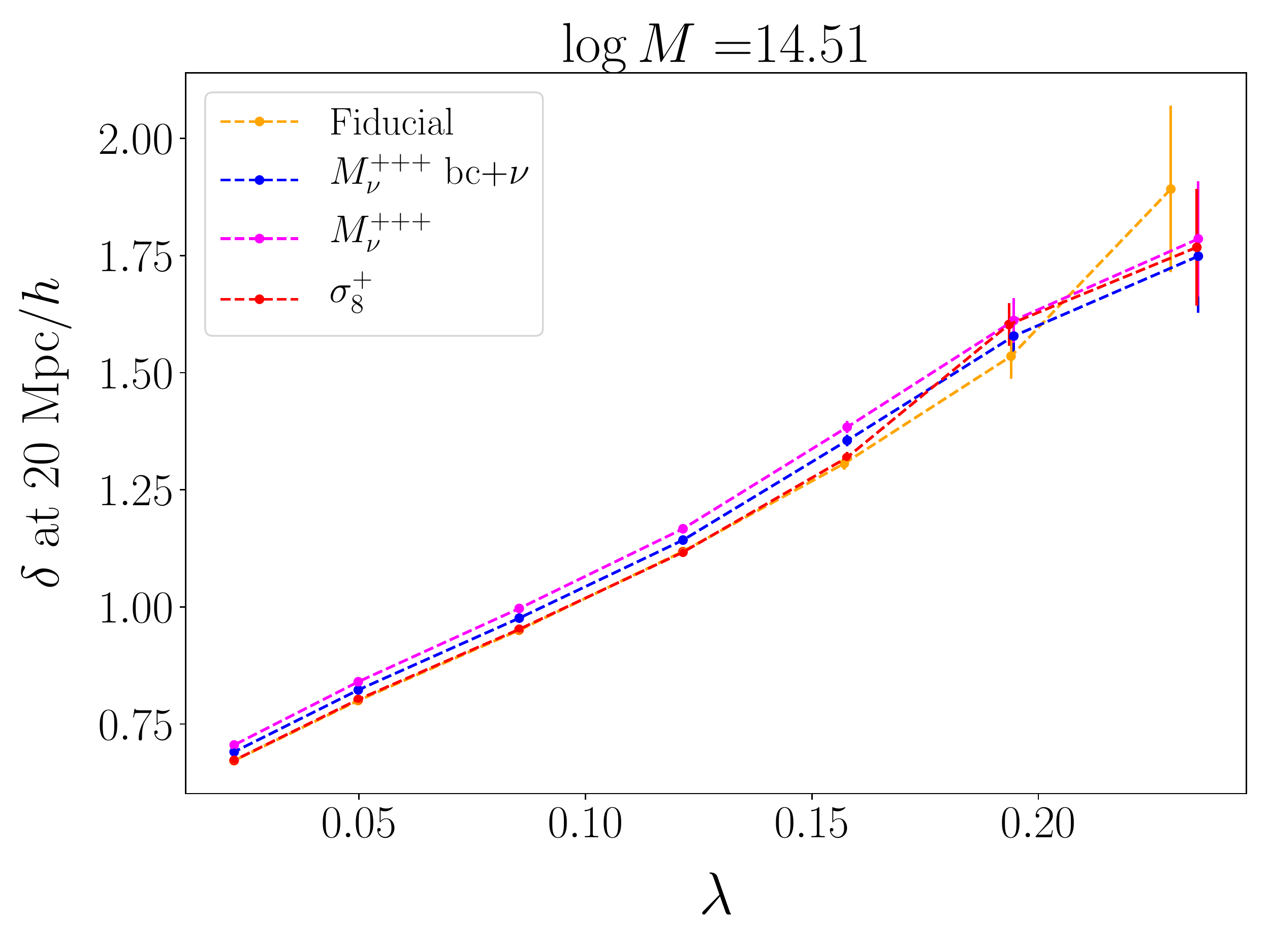}
\includegraphics[scale=0.275]{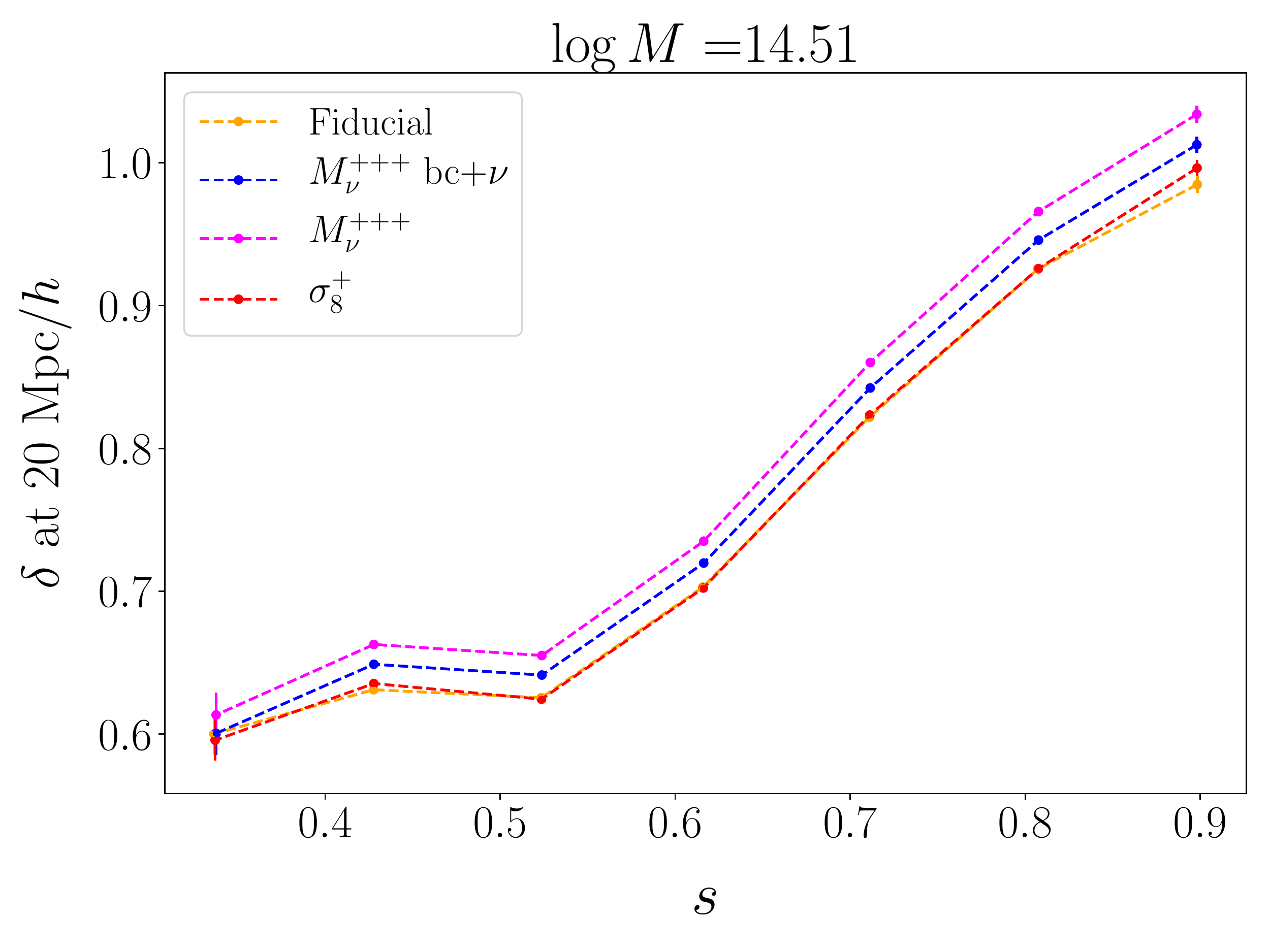}
\caption{Overdensity $\delta$ in a sphere of radius $R=20 \Mpch$ around halos as a function of halo properties at fixed mass $\log M = 14.51$ for different cosmologies indicated by the color coding, as on \reffig{propsofM}. The top left panel shows $\delta$ as a function of concentration, the top right one as a function of spin, and the bottom one as a function of ellipticity (more spherical halos are on the right). This time halos include neutrinos by default in the $M_\nu^{+++}$ cosmology. The magenta curve shows results when considering only the baryon-CDM fluid to define the overdensity, while the blue one shows result when considering the total matter field. We see that $\sigma_8$ does not affect this quantity while neutrinos clearly enhance the overdensity in a seemingly constant manner over the range of value taken by each property. The fact that the overdensity including neutrinos (blue lines) is lower than without (magenta lines) is simply due to the fact that the background density is higher when neutrinos are included. We have checked that other mass bins or radius exhibit similar trends.}
\label{fig:deltaofp}
\end{figure}

We now turn to the impact of neutrinos on the large scale density of halos, defined as the overdensity in a sphere of radius $R=20 \Mpch$ around the halo center, as a function of halo properties. This is motivated by the fact that halo assembly bias has now been shown to be related to halo environment and position in the cosmic web \cite{Borzyszkowski:2016, Musso:2017, Ramakrishnan:2019}. For example, high mass halos (that dominate their environment) with a lower concentration are expected to have a higher large-scale density on average since they have accreted less material. The average (or stacked) large-scale density of halos in turn corresponds to the cross correlation of the matter with the halo field, and is hence linked to the linear bias (higher environmental density gives rise to a higher bias). In this picture it is then clear why less concentrated objects are more biased, as can be seen on the top row of \reffig{bofp}. Since we do not see any strong difference between assembly bias in the fiducial and neutrino cosmologies, we do not expect them to affect the large scale density of halos in a property-dependent way. More precisely, while neutrinos might affect the overall overdensity around halos, we expect this effect to be constant over the range of value taken by a given property.

\begin{figure}
\centering
\includegraphics[scale=0.275]{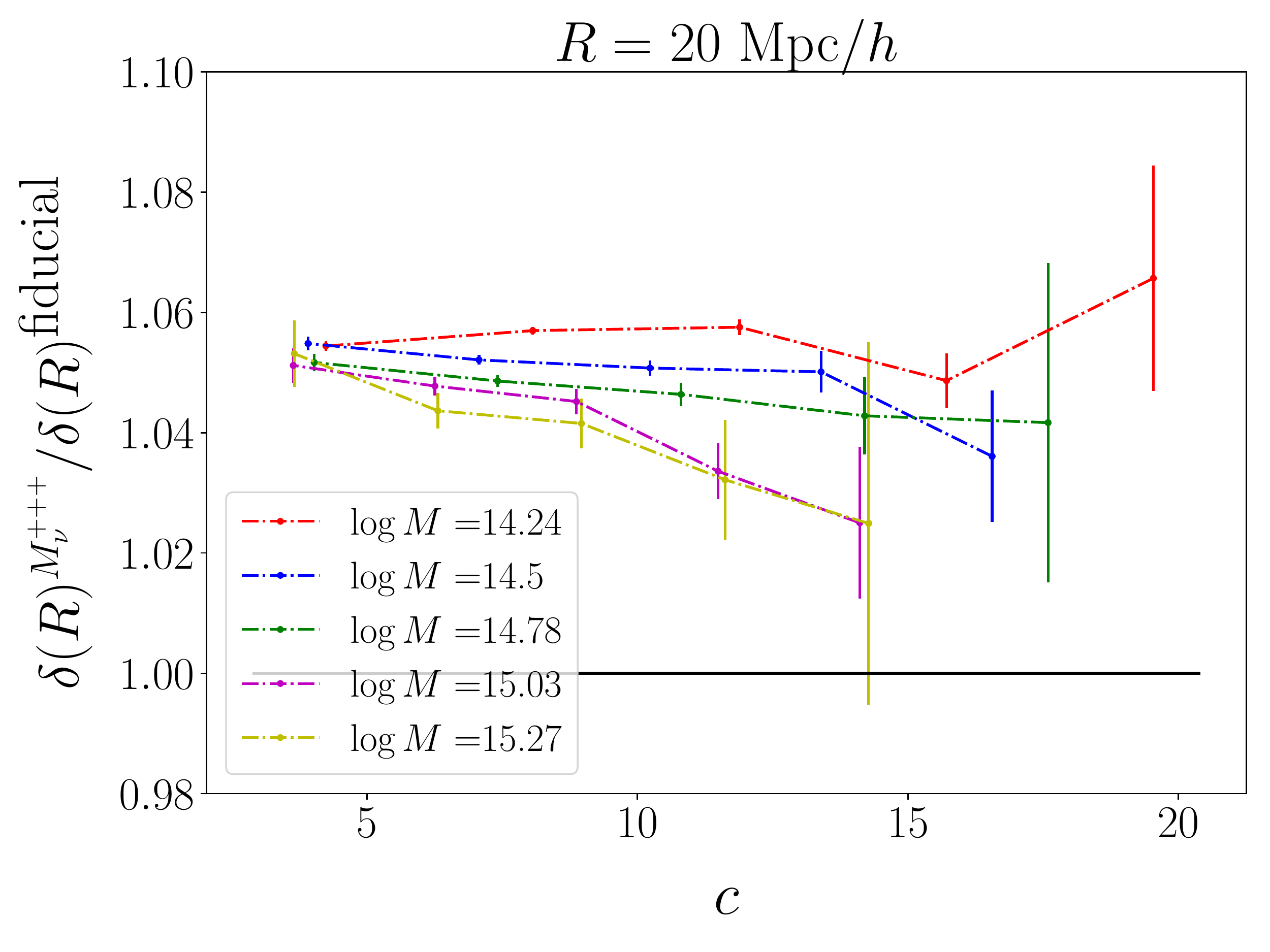}
\includegraphics[scale=0.275]{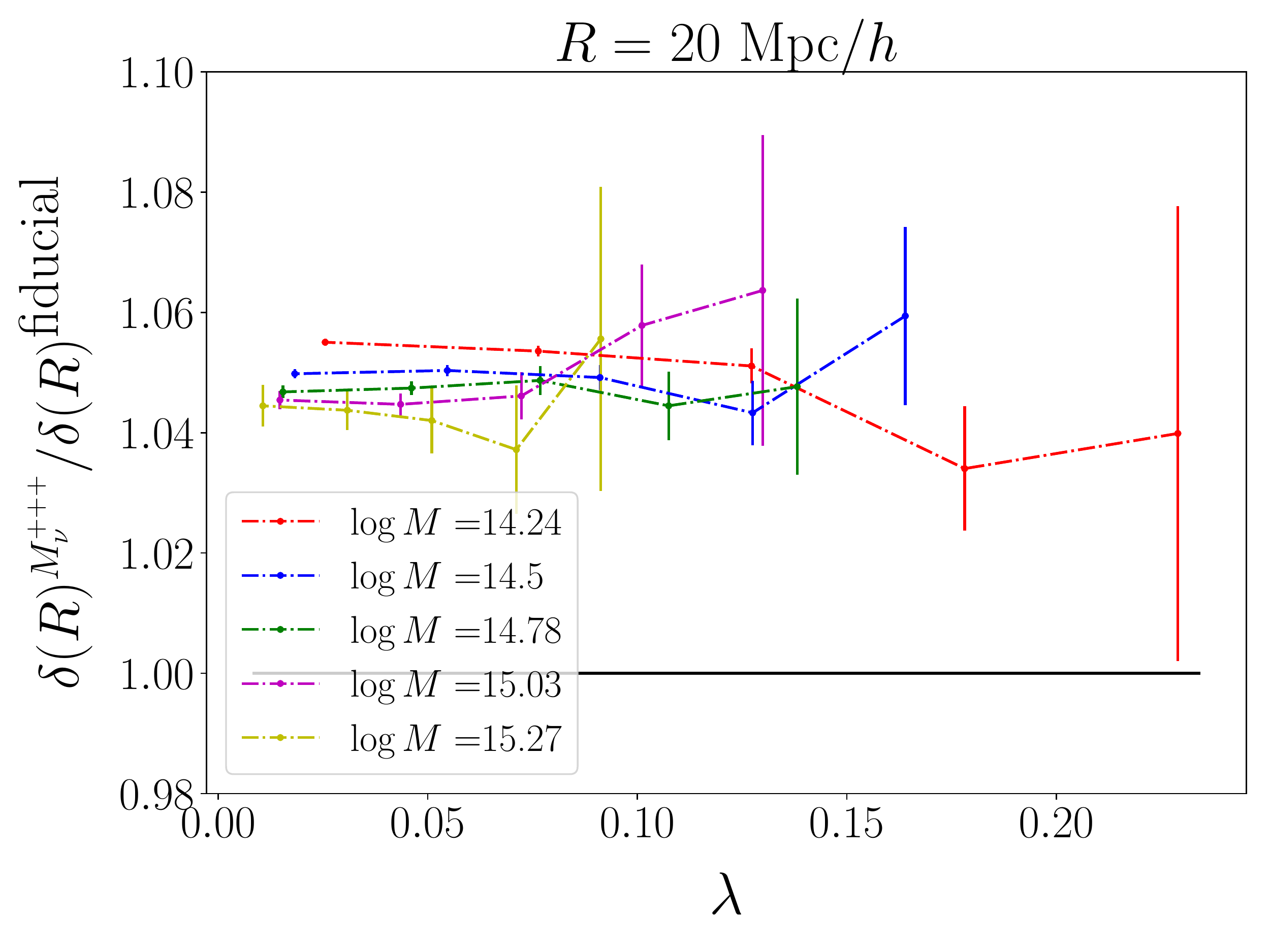}
\includegraphics[scale=0.275]{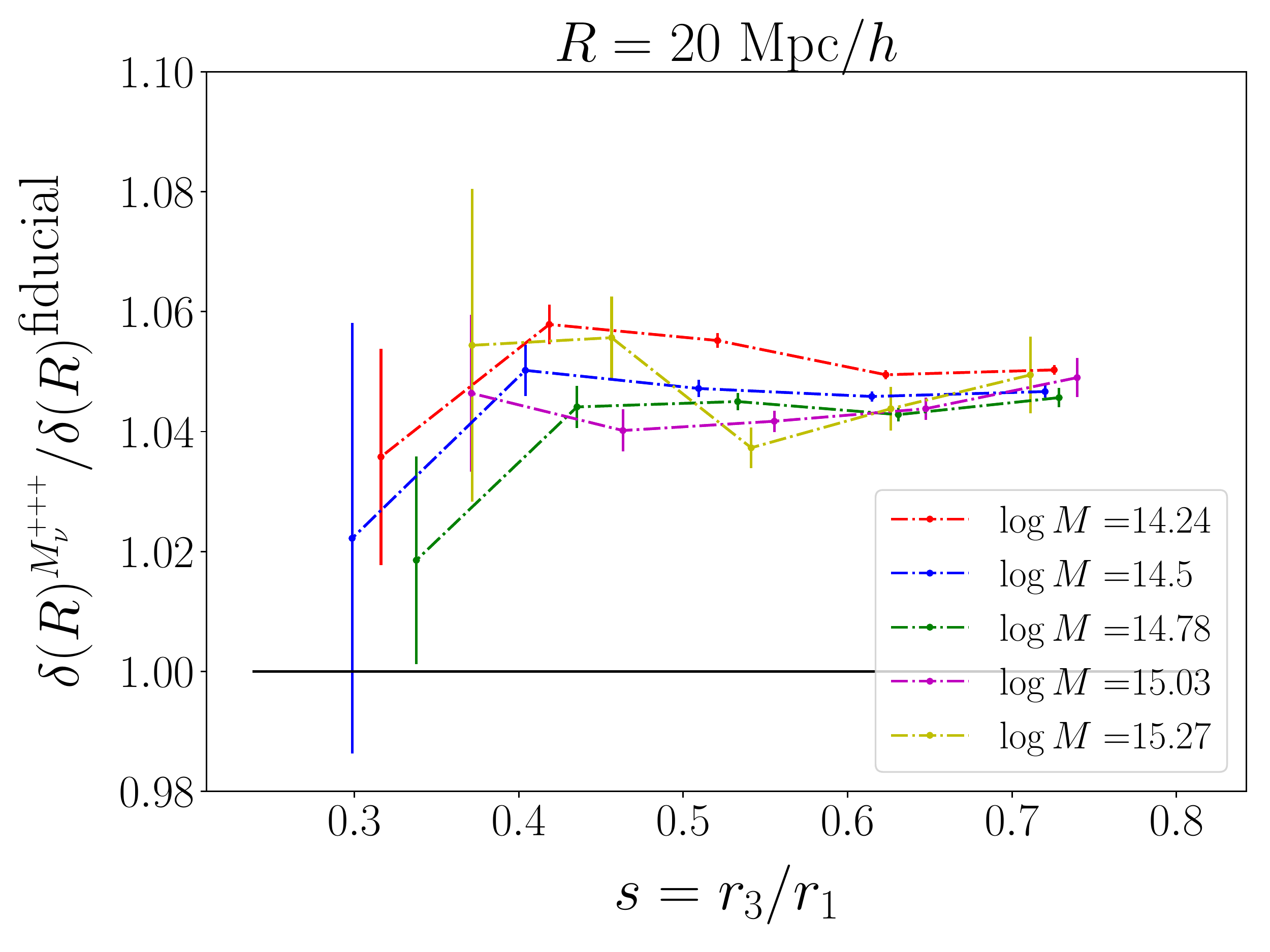}
\caption{Ratio of the overdensity $\delta$ at $R=20 \Mpch$ between the $M_\nu^{+++}$ and fiducial cosmology as a function of halo properties ($c$ on the top left panel, $\lambda$ on the top right, and $s$ on the bottom panel). The color coding indicates the halo mass and is the same as on figures in \refsec{assemblynu}. We do not include neutrinos in the large-scale overdensity computation (i.e. magenta curves on \reffig{deltaofp}) but they are included in halos. Including neutrinos to compute $\d$ does not change the shape of the curves but simply the overall amplitude of the ratio, as can be seen from \reffig{deltaofp}. The ratios are close to constant (within errorbars) for each property and at each mass, which is consistent with the results presented on \reffig{ratios}. See text for more details.}
\label{fig:ratiodMnupppnn}
\end{figure}

\refFig{deltaofp} shows the results for the mean overdensity in a radius of $20 \Mpch$ around halos of $\log M = 14.51$ as a function halo properties. The color coding is the same as in \reffig{propsofM} but we do not consider the $M_\nu^{++}$ cosmology. We show only results for one mass bin for shortness and clarity but we checked that other mass bin behave similarly. We have also checked that the results do not depend on the specific value of $R$ by looking at the overdensity in spheres of radii $5$, $10$, and $30 \Mpch$ and finding similar trends.  Regarding the general trends common to all cosmologies, on the top left panel we observe that the overdensity decreases as concentration increases which is coherent with the picture that we explained before. The environment density increases with both spin and sphericity of halos as was found in e.g. \cite{Lee:2016} for high-mass halos. 

We see that $\sigma_8$ does not affect these quantities while neutrinos have a visible effect that seems constant over the range of value taken by each property. This time halos include neutrinos by default in the $M_\nu^{+++}$ cosmology. The magenta curve shows results when considering only the baryon-CDM fluid to define the overdensity, while the blue one shows result when considering the total matter field. A striking point is the fact that including neutrinos or not has an important impact on the environmental density, which is clearly higher when using the baryon-CDM fluid only. This is simply because the background density with respect to which the local overdensity is computed increases when neutrinos are included, while they have a negligible impact on the local density of halos on such small scales. Hence the overdensity diminishes.

Again, in order to be more quantitative, we show the ratio of the magenta curves (i.e. $M_\nu^{+++}$ with the overdensity defined in the baryon-CDM fluid) to the fiducial cases (orange) on \reffig{ratiodMnupppnn}. This time the color coding indicates halo mass and is the same as on \reffig{ratios}. The overall amplitude of these ratio is slightly mass dependent but is of no particular interest for our case. The most important result of these figures is that these ratios are constant over the range of mass and halo properties considered, within the errorbars. Since, as explained before, these ratios are linked to the ratios of assembly bias in neutrino and fiducial cosmologies, we expected this result. In particular the dependence of each ratio on each property is of maximum a couple percent, and the trends go in the same way as on \reffig{ratios} (notice for example the small mass dependence of these ratios as a function of concentration on the top left panel of \reffig{ratiodMnupppnn} with more massive and more concentrated halos having a ratio slightly lower). Hence, as expected, neutrinos affect the large-scale density of dark matter halos in the same way and at the same level as their assembly bias.

\section{Conclusions}
\label{sec:concl} 

We have presented new measurements of the assembly bias of dark matter halos using the Quijote simulations which incorporate the effect of massive neutrinos. We studied their impact on the linear bias as a function of concentration, spin parameter, and ellipticity. Our main result is that \textit{neutrinos do no affect assembly bias of cluster sized halos} for any of the property considered and any of the neutrino mass considered. This result is independent of including neutrinos in halos or not. We do detect some minuscule trends that are not statistically relevant but do not try to confirm them by, e.g. adding more simulation volume, since the volume we use here ($500 ({\rm Gpc}/h)^3$) is already far superior to any current or future survey volume.

While our findings imply that constraining neutrino masses using assembly bias looks challenging, they do have the advantage of simplifying the next generation of galaxy surveys aiming at percent level precision cosmology. Indeed if neutrinos were shown to affect assembly bias in a measurable way, this effect would have to be included when going beyond the standard $\LCDM$ model of cosmology. Furthermore, any cosmological constraint made using assembly bias in a $\LCDM$ Universe will now immediately hold as well in neutrinos cosmologies.

We emphasis that, in the present work, we ruled out neutrino effects on assembly bias for cluster-sized objects, which are known to generally reside at the intersection between filaments in the cosmic web. There is a population of low mass halos ($10^{12} M_\odot/h$ and lower) residing in filaments for which assembly bias is known to be important, and to be generated via other mechanisms than for cluster-sized halos (e.g. \cite{Hahn:2008yh}). Due to the mass resolution of the Quijote simulations, we cannot probe this mass regime.

We further investigated the impact of neutrinos on halo properties and large-scale environment (i.e. the overdensity in a sphere of radius $R=20\Mpch$). To our knowledge this is the first time that the effect of neutrinos on halo properties is studied in details. Again we did not find any strong impact of neutrinos on any of these quantities, apart from a constant positive shift in the large-scale density. This was expected since any change in halo properties should lead to a change in assembly bias via an Eddington-like bias, while any property dependent change in the large-scale density would be related to a signal in assembly bias. This last point results from the fact that the stacked large-scale overdensity of matter around halo is equal to the halo-matter correlation function which in turn determines the linear bias of halos. Hence our findings for the impact of neutrinos on halo properties and environment are consistent with our results for assembly bias in neutrino cosmologies.

\acknowledgments{We thank Marcello Musso and Tiago Castro for interesting discussions and useful comments. We also thank Alexander Knebe for providing details on the AHF halo finder. TL thanks CCA for their hospitality during part of this work. FVN acknowledge funding from the WFIRST program through NNG26PJ30C and NNN12AA01C. MV is supported by INFN PD51-INDARK grant and by a grant from the agreement ASI-INAF n.2017-14-H.0.}

\appendix

\section{Effect of including neutrinos in halos}
\label{app:nuhalo}

As we explained in the main text, we do not include neutrinos during the halo finding procedure. In other term we assume that neutrinos impact the clustering of matter but do not cluster themselves. This goes along the same line as when considering only the baryon-CDM fluid power spectrum when computing the bias. In this section, we investigate whether including neutrinos in halos impacts our results for assembly bias. We focus on simulations with the highest neutrino mass $M_\nu^{+++}$ since this is where the difference should be maximum.

\begin{figure}
\centering
\includegraphics[scale=0.275]{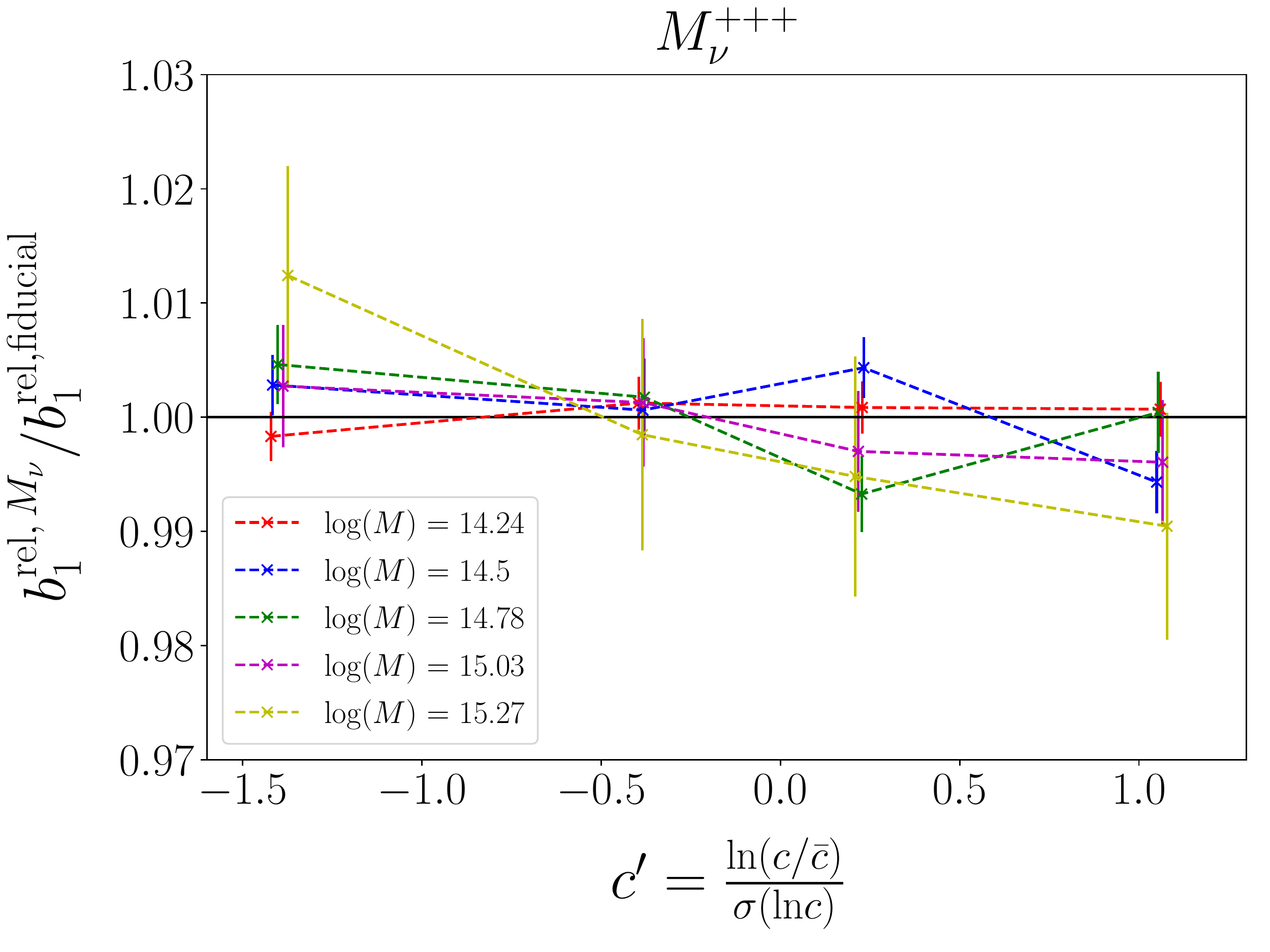}
\includegraphics[scale=0.275]{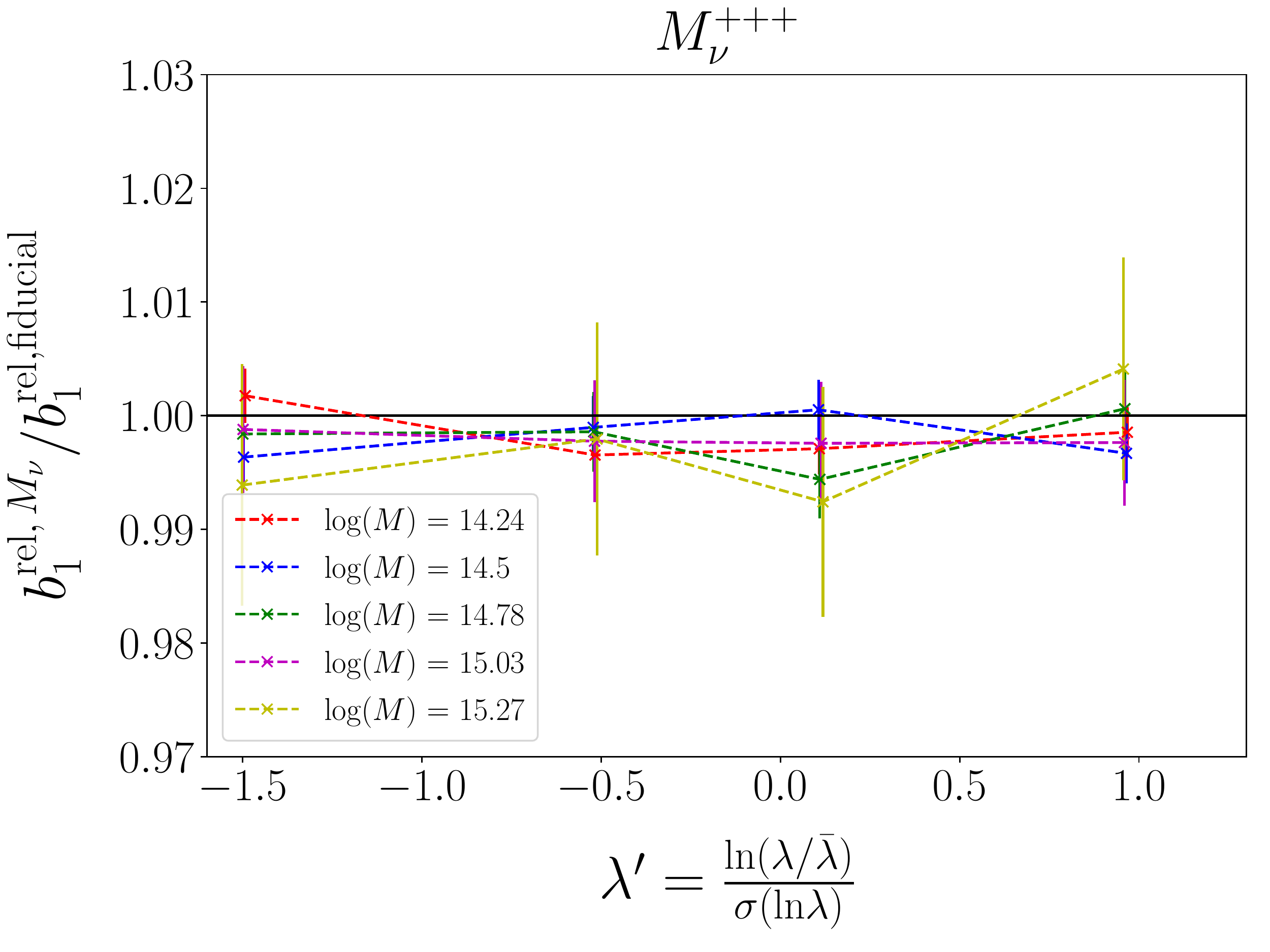}
\includegraphics[scale=0.275]{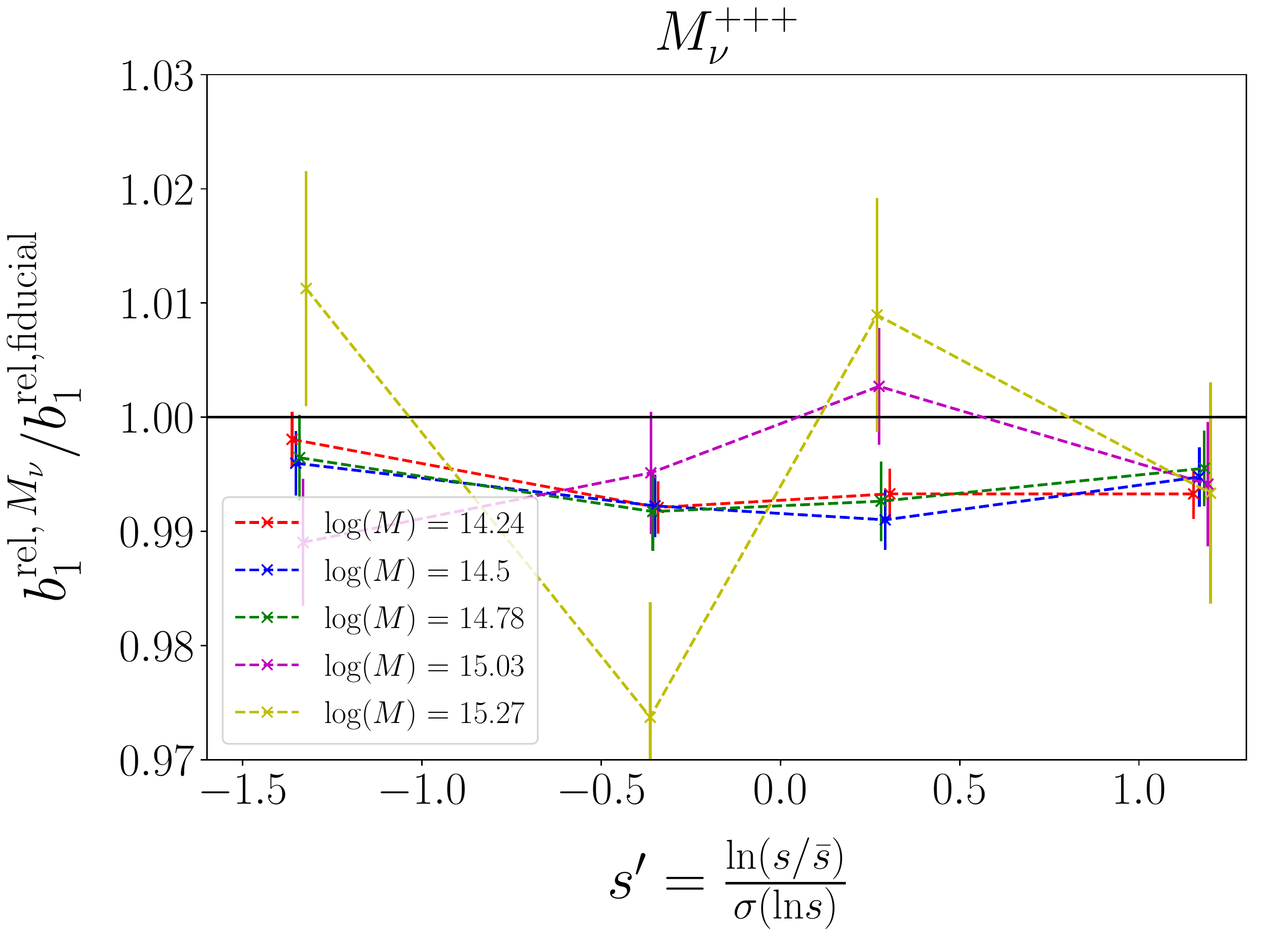}
\caption{Same as \reffig{ratios} but as measured when including neutrinos in the halo finding procedure. The top left panel shows results for halo bias as a function of concentration, the top right one for halo spin, and the bottom one for halo ellipticity (spherical halos have $s'=1$). The color coding is the same as on \reffig{ratios}. As can be seen by comparing these results with the ones in the main text, including neutrinos in the halo definition has no visible effect on their impact on assembly bias.}
\label{fig:ratiownu}
\end{figure} 

\refFig{ratiownu} shows the same ratios as \reffig{ratios}, but for halos found in the baryon-CDM+ neutrinos fluid. The panels from top left in clockwise order show the bias as a function of concentration, spin and sphericity, with the same color coding for halo mass as in the main text. We see no visible difference between these plots and the ones presented in \refsec{assemblynu}. Hence including neutrinos in the halo definition has virtually no effect on our results. This is expected since neutrinos have a high velocity dispersion and are not expected to cluster substantially on halo scales. In addition, AHF removes unbound particles which would diminish their abundance in halos further.

To further assess the importance of including neutrinos in the halo finding procedure we looked at the halo catalogs themselves and found a subpercent effect on the halo mass function, as was previously known \citep{Castorina:2013, Costanzi_2013}. We also found that including neutrinos in halos has a subpercent effect on the inferred mass and all halo properties. This was however an important point to check since halo properties could a priori be different when using the baryon+CDM or baryon-CDM+neutrino fluid. Furthermore assigning, e.g. an artificially too low concentration to halos of a given mass would tend to enhance the bias leading to an Eddington-like bias in the case of mass. This is however not the observed trend on \reffig{ratios} and \reffig{ratios2}. 

We hence conclude that including neutrinos or not in halos is of no importance for our study. However, in order to stay consistent and only assess the effect of neutrinos on the clustering without assuming that they cluster themselves we focus our study on halos made of the baryon-CDM fluid only (bc halos in the main text).

\section{Additional neutrinos masses}
\label{app:addnumass}

\begin{figure}
\centering
\includegraphics[scale=0.275]{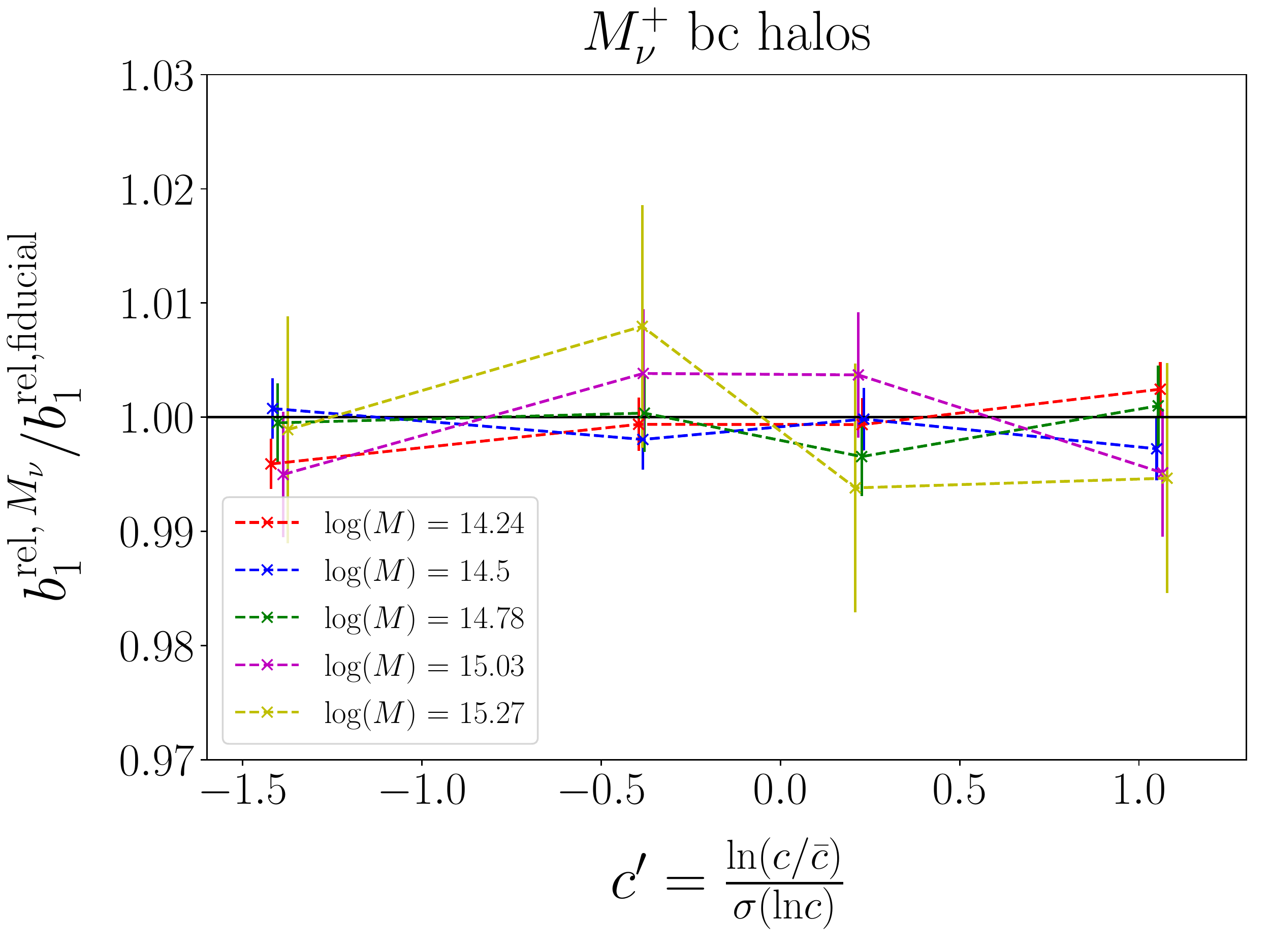}
\includegraphics[scale=0.275]{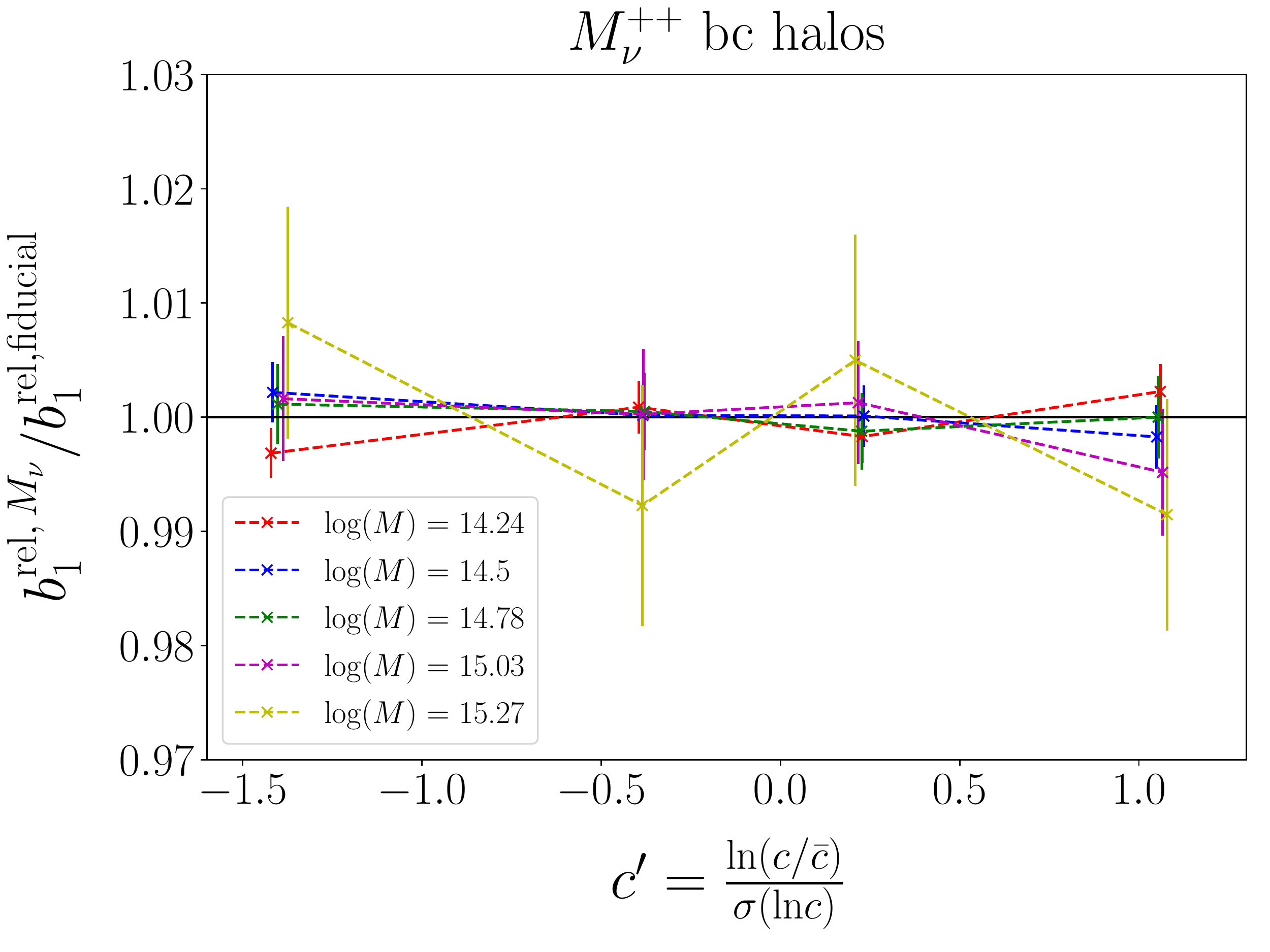}
\includegraphics[scale=0.275]{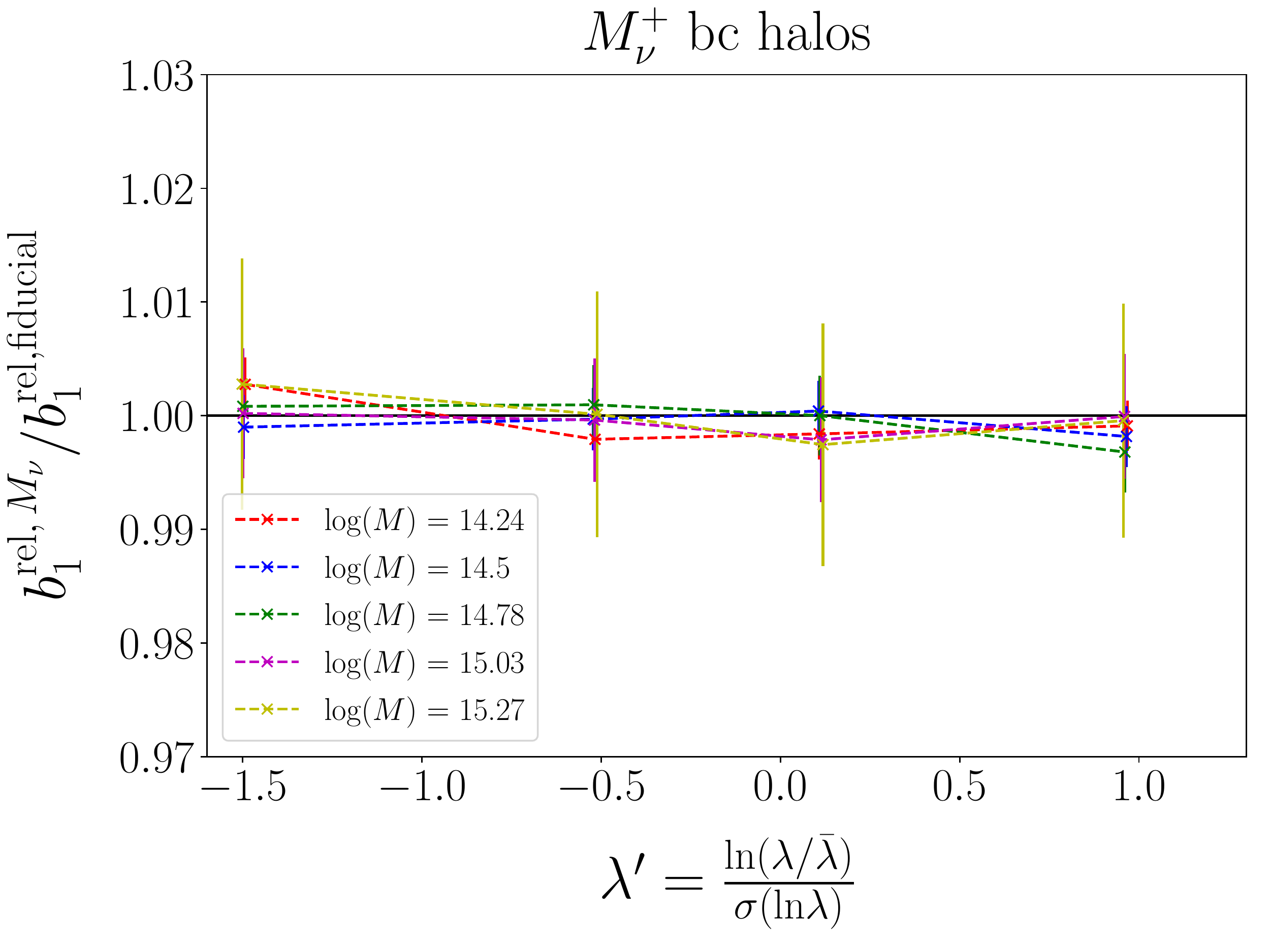}
\includegraphics[scale=0.275]{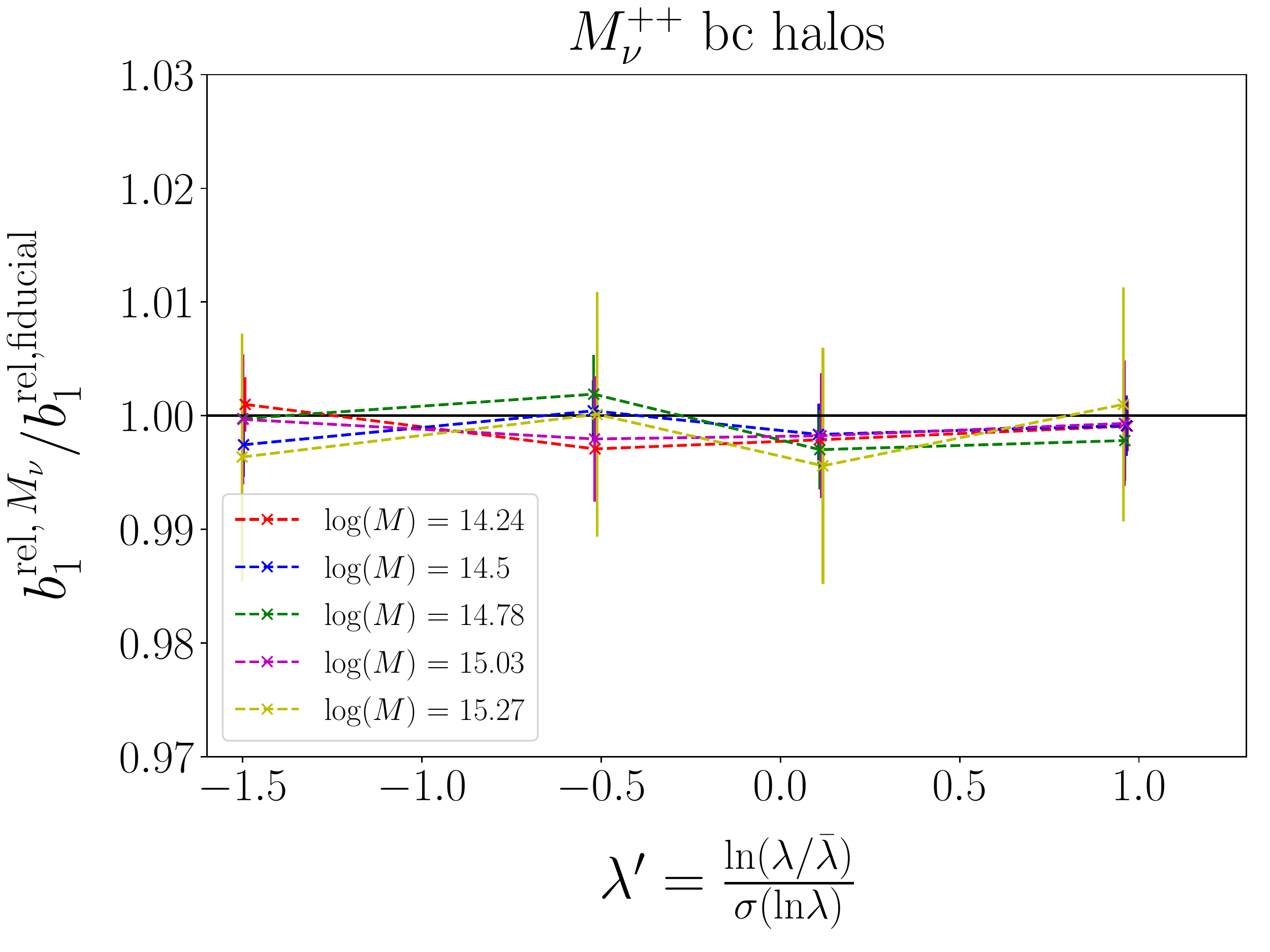}
\includegraphics[scale=0.275]{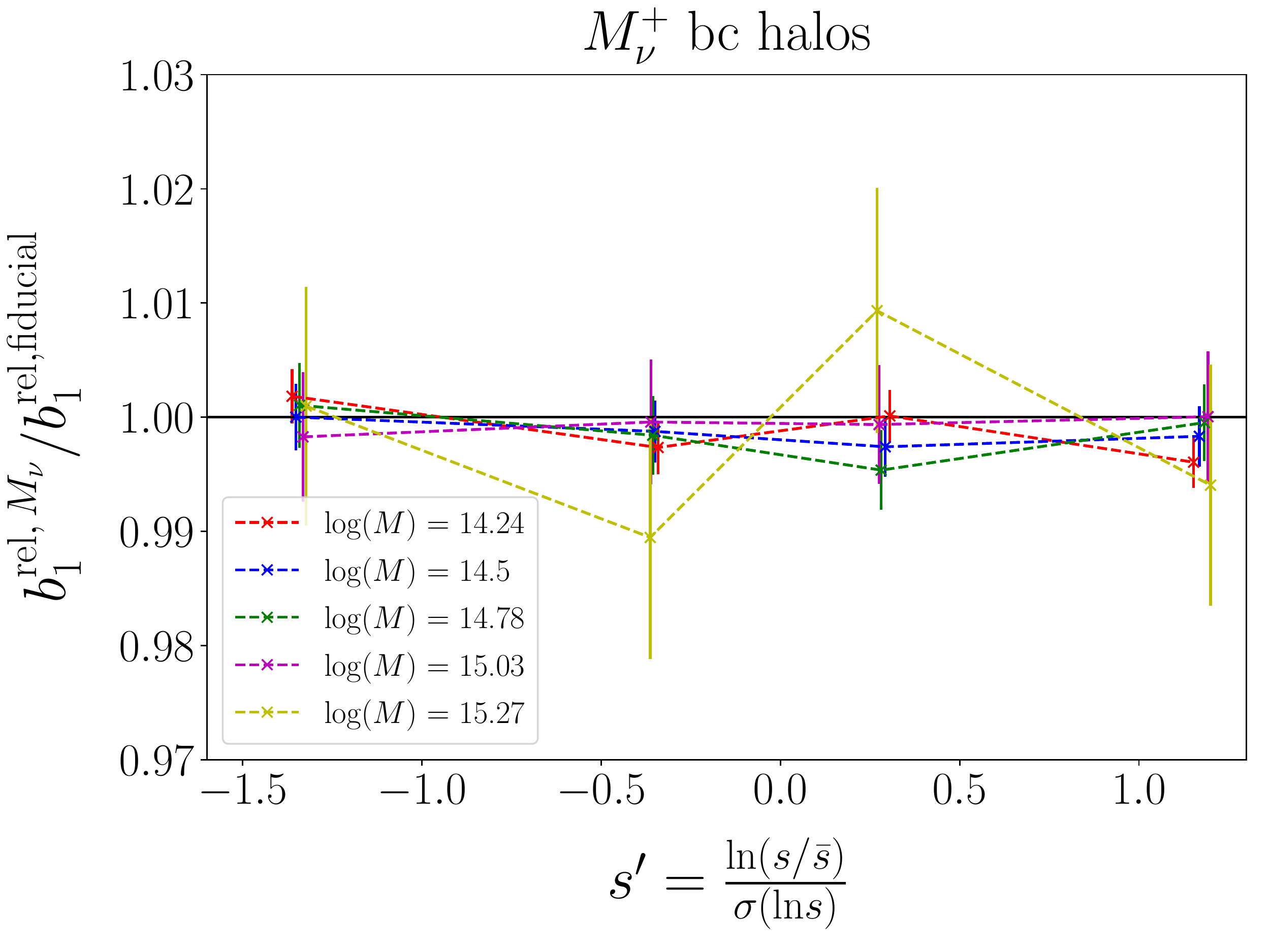}
\includegraphics[scale=0.275]{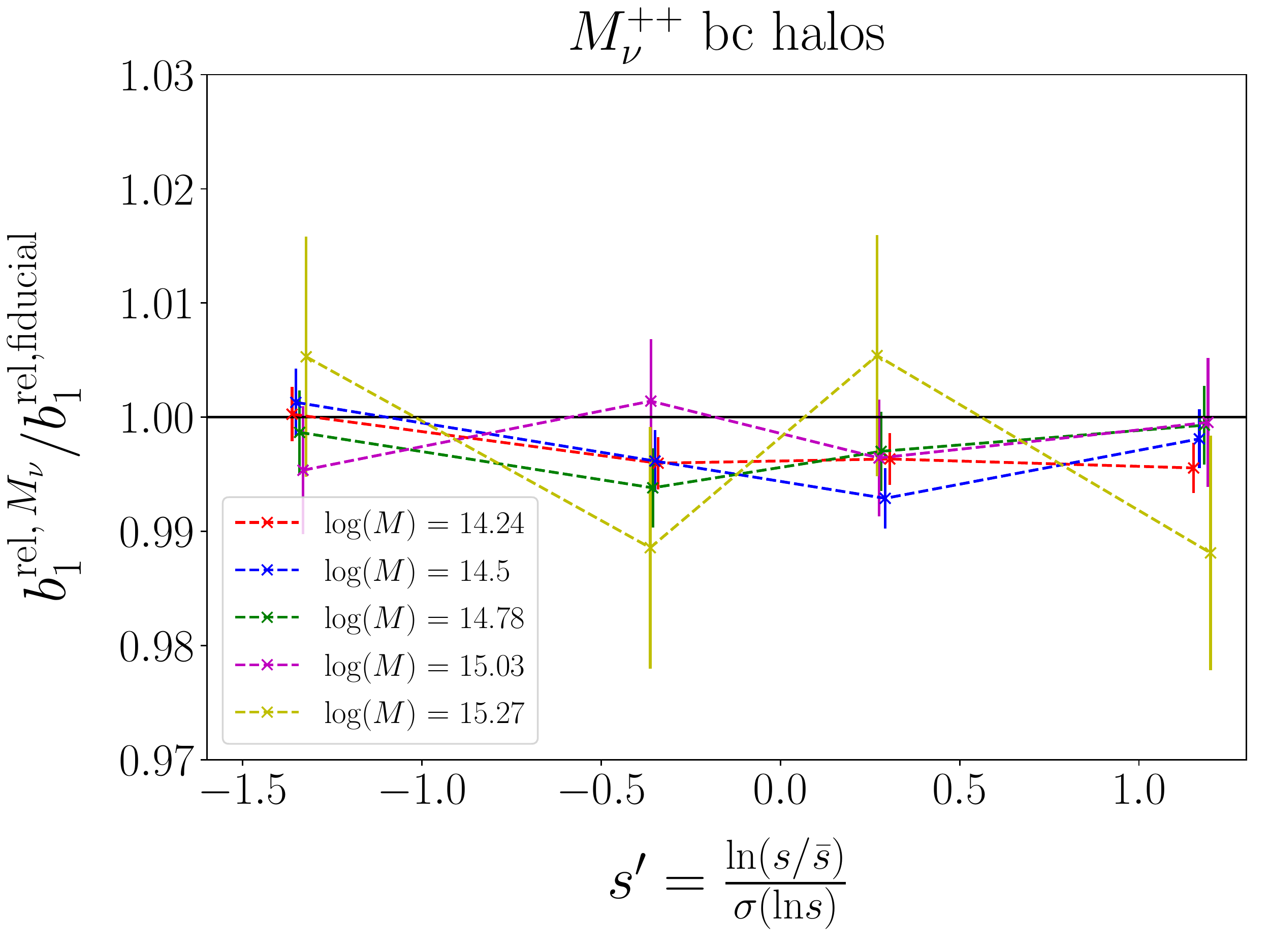}
\caption{Same as \reffig{ratios} but for smaller neutrino masses. The panels of the left column show results for simulations with neutrinos of a total mass of 0.1 eV while the right ones show those for the $M_\nu^{++}$ cosmology, i.e. $M_\nu=0.2$ eV. Halos are found using the baryon-CDM fluid only. As before no effect of neutrinos on the bias as a function of any of the properties can be detected, which is expected since the maximum effect should be for the highest neutrino mass, shown in the main text.}
\label{fig:ratios2}
\end{figure}

For the sake of completeness we show in this appendix the same results as in \refsec{results} but for neutrinos masses of 0.1 and 0.2 eV. The results are presented on \reffig{ratios2}. The left column presents results in the $M_\nu^+$ cosmology, while the right one presents those in the $M_\nu^{++}$ one. The first row shows results for concentration, the middle one for spin, and the bottom one for sphericity. The color coding is the same as on \reffig{ratios}.  

As we explained before, we expect the effect of neutrinos to be maximal for the highest mass, which is confirmed by \reffig{ratios2}. Indeed all curves are consistent with one to an even higher statistical significance than those shown in the main text. The neutrinos mass considered in this appendix are however more reasonable (in particular 0.1 eV is not yet ruled out) which is why we show their corresponding results. 

The results presented in this appendix further confirm that neutrinos are not expected to have any measurable effect on assembly bias. Hence it will be challenging to constraint neutrino masses in this way. However this also means that it will be possible to safely ignore this effect in analysis of future survey data, in case assembly bias is proven to affect also galaxy and galaxy clusters, and when going to cosmologies beyond the standard $\LCDM$ model. 

\section{Impact of $\sigma_8$ on assembly bias}
\label{app:sigma8}

The simulations including massive neutrinos keep $\sigma_8$ fixed with respect to the fiducial cosmology. However this is $\sigma_8$ for the total matter fluid, and not for the baryon-CDM component only, which could be the more relevant quantity. This means that the amplitude of the primordial fluctuations is not kept constant. Basically, if we fix $A_s$, the neutrino cosmology will have a lower $\sigma_8$, and therefore halos of the same mass will form later. Since halo properties such as concentration, spin and shape can in general depend on halo formation time, neutrinos could leave an imprint in halo assembly bias, but just due to the change in $\sigma_8$ induced by the numerical implementation. We must hence check the impact of a change in $\sigma_8$ of the baryon-CDM component on assembly bias. 

\begin{figure}
\centering
\includegraphics[scale=0.275]{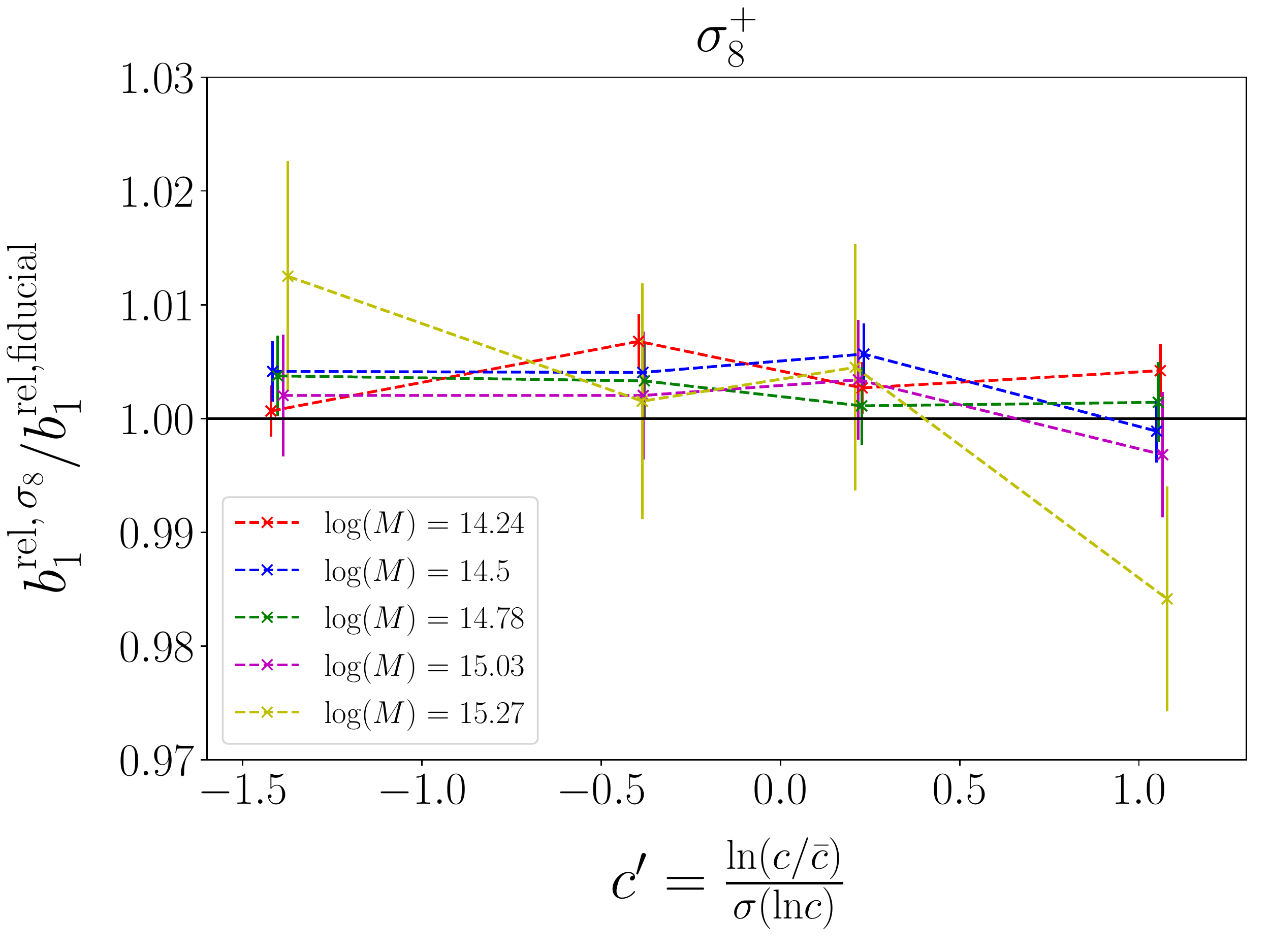}
\includegraphics[scale=0.275]{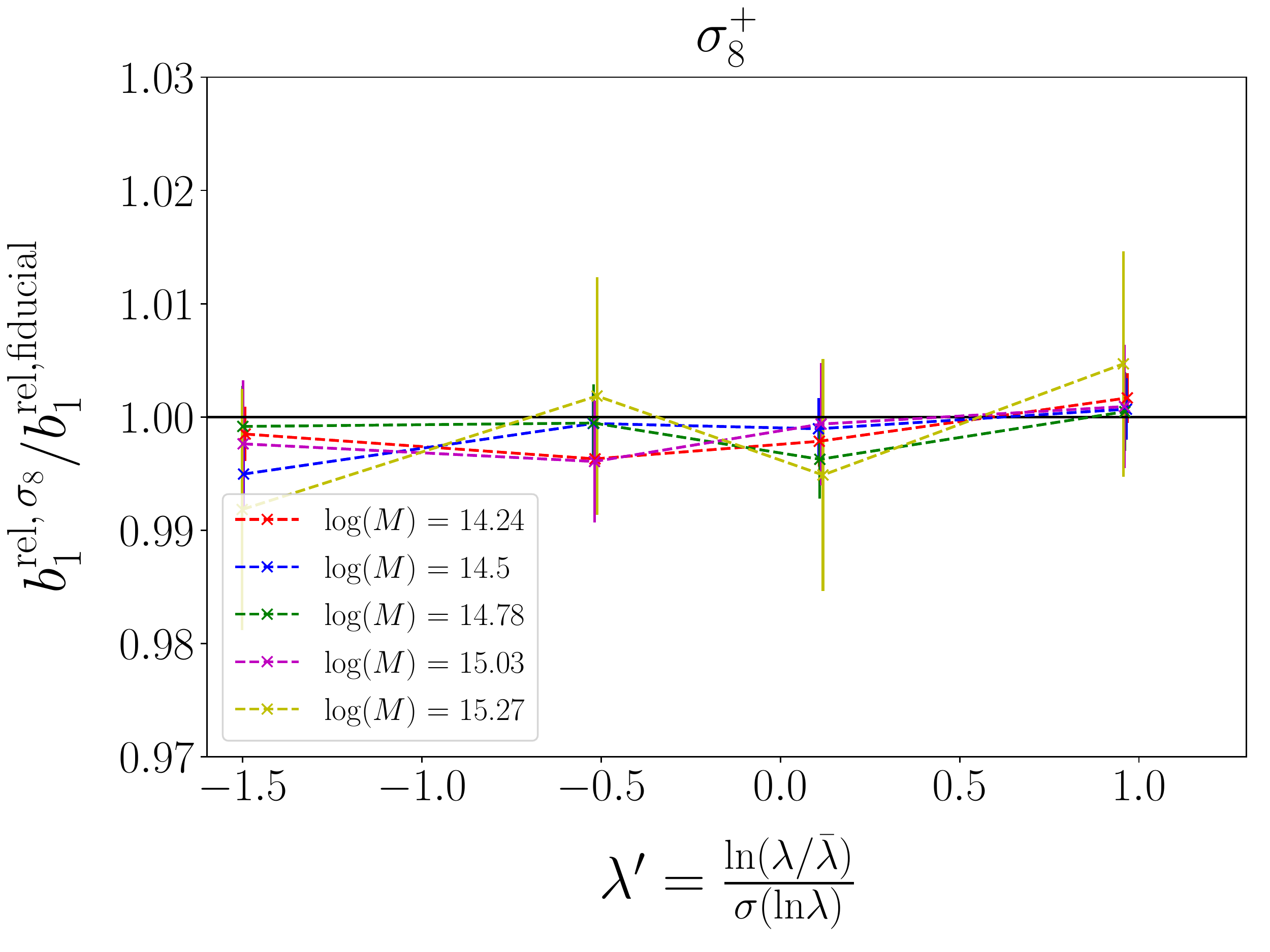}
\includegraphics[scale=0.275]{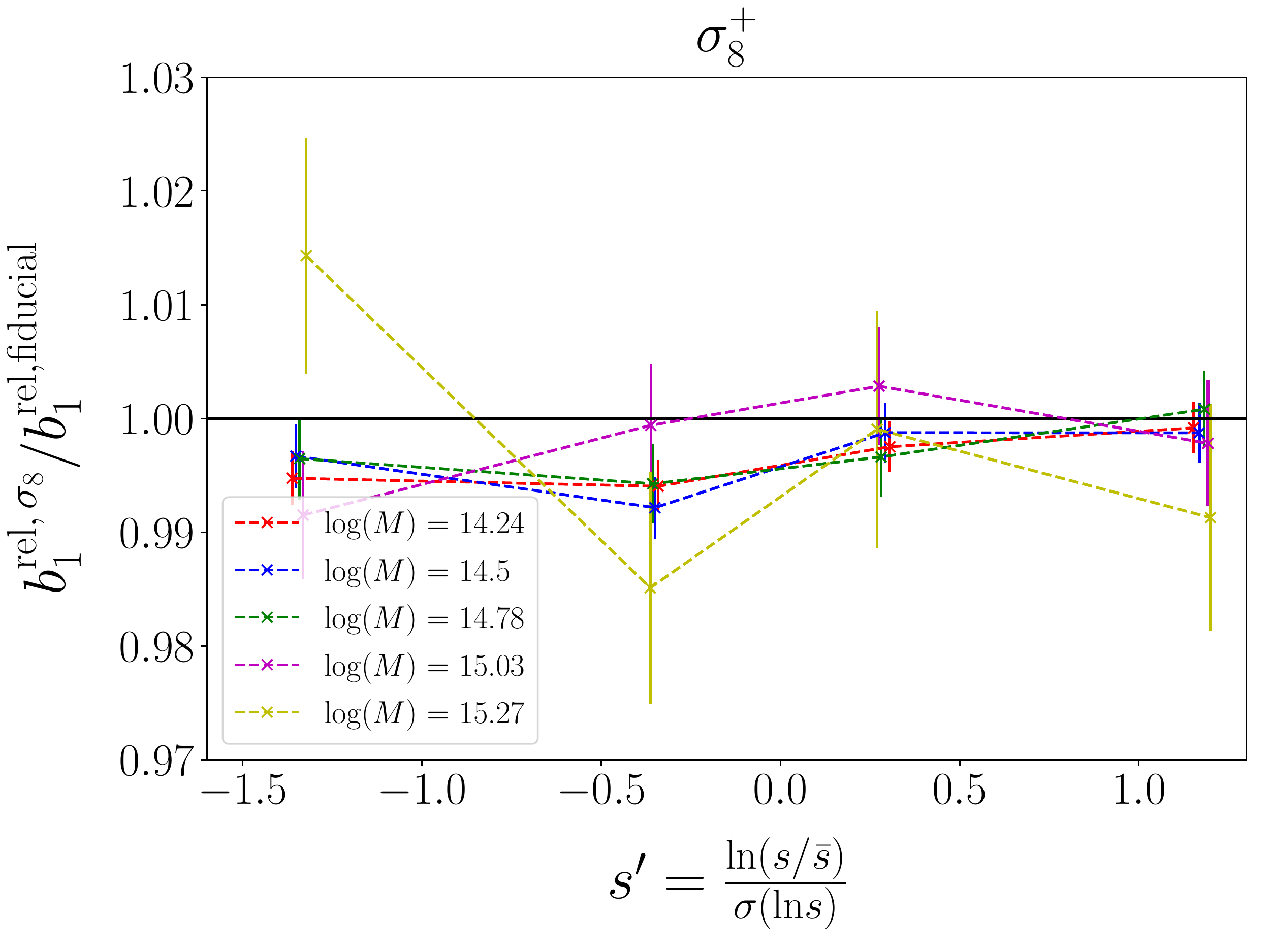}
\caption{Ratio of assembly bias in $b_1$ measured in simulations with enhanced $\sigma_8$ to the one measured in the fiducial cosmology as a function of halo properties for several mass bins (indicated by the color code). The errorbars show the $1\sigma$ error obtained in the same way as for \reffig{ratios}. The top left panel shows the bias as a function of concentration, the top right one as a function of spin parameter, and the bottom one as a function of ellipticity. While the effect of enhancing $\sigma_8$ is negligible on halo assembly bias, as can be seen, all trends go in the same direction as the ones presented on \reffig{ratios}, which further diminishes the impact of neutrinos.}
\label{fig:ratiosigma8}
\end{figure}

For this we use an additional set of gravity-only simulations without neutrinos but with an enhanced $\sigma_8$. This mimics the change of $\sigma_8$ due to neutrinos. In particular the adopted value of $\sigma_8=0.849$ in the $\sigma_8^+$ cosmology is close to the one of baryon-CDM in the $M_\nu^{++}$ cosmology ($\sigma_{8,c}=0.846$). We then reproduce the same pipeline as in the case of massive neutrinos to see if we observe similar trends. The results are presented on \reffig{ratiosigma8}. Each panel shows the ratio $b_1^{\rm rel, \, \sigma_8^+}/b_1^{\rm rel, \, fiducial}$ as a function of a halo property at fixed mass (indicated by the color coding).

Halo assembly bias is only very little affected by a change in $\sigma_8$. All curves on \reffig{ratiosigma8} are consistent with 1 within the errorbars. By eye we can however see some very small trends which are not statistically significant but that go in the same direction as the ones we described on \reffig{ratios}. This further diminishes the possibility to detect any impact of neutrinos on assembly bias since our results seem to point to a degeneracy between results in neutrino cosmologies and the enhanced $\sigma_8$ one.

\FloatBarrier
\bibliography{references}
\end{document}